\documentclass[12pt]{article}

\usepackage{theorem,amssymb,amsmath,amsbsy,latexsym}
\usepackage{graphicx}

\newcommand{\file}{eps}
\newcommand{\HttpV}{H_{tt'V}}
\newcommand{\HLutt}{\tilde{H}_a}
\renewcommand{\Re}{\textrm{Re}}
\renewcommand{\Im}{\textrm{Im}}
\newcommand{\EE}{\varepsilon} 
\newcommand{\mh}{\boldsymbol{h}}
\newcommand{\mgamma}{\boldsymbol{\gamma}}
\newcommand{\id}{\boldsymbol{1}}
\newcommand{\OmHF}{\Omega_{\mathrm{HF}}} 
\newcommand{\HFG}{\mathrm{HF}} 
\newcommand{\uno}{{u}} 
\newcommand{\ttpV}{\mbox{$t$-$t'$-$V$ }}
\newcommand{\cN}{\mathcal{N}}
\newcommand{\cT}{\mathcal{T}}
\newcommand{\cP}{\mathcal{P}}
\newcommand{\cR}{\mathcal{R}}

\newcommand{\cC}{\mathcal{C}}
\newcommand{\cE}{{\cal E}}  
\newcommand{\CDW}{\mathrm{CDW}}
\newcommand{\N}{\mathrm{N}} 
 
\newcommand{\Heff}{H_{\textrm{eff}}} 
\newcommand{\geff}{g_{\textrm{eff}}}
\newcommand{\ELref}[1]{(I.{#1})}

\newcommand{\tQ}{Q_0}
\newcommand{\QQPi}{\mbox{$\frac{2Q}{\pi}$}} 
 
\newcommand{\tQQPi}{\mbox{$\frac{2\tQ}{\pi}$}} 
\newcommand{\pdag}{^{\phantom\dag}} 
 
\newcommand{\half}{\mbox{$\frac12$}}
\newcommand{\third}{\mbox{$\frac13$}}
\newcommand{\ee}{\,{\rm e}}
\newcommand{\BZ}{\mathrm{BZ}}
\newcommand{\Z}{{\mathbb Z}}

\newcommand{\vac}{{|\mathrm{vac}\rangle}}
\newcommand{\mbf}[1]{{\boldsymbol {#1} }}
\newcommand{\ii}{{\rm i}}
\newcommand{\define}{\stackrel{\mbox{{\tiny def}}}{=}}
\newcommand{\vx}{{\bf x}}

\newcommand{\vn}{{\bf n}}
\newcommand{\vk}{{\bf k}}
\newcommand{\vQ}{{\bf Q}}
\newcommand{\vzero}{{\mbf 0}}
\newcommand{\ve}{{\bf e}}
\newcommand{\vp}{{\bf p}}
\newcommand{\Ref}[1]{(\ref{#1})}
\newcommand{\eps}{\epsilon}
\newcommand{\tPiL}{\mbox{$\frac{2\pi}{L}$}}

\begin{document}
\begin{flushright}
April 8, 2010
\end{flushright}
\vspace{.4cm}

\begin{center}
  \vspace{1 cm}
 {\Large \bf Partially gapped fermions in 2D}\\
  [2mm]
  {\bf Jonas de Woul and Edwin Langmann} \\
  [2mm]
  {\it Theoretical Physics, KTH\\
    SE-10691 Stockholm, Sweden}\\
  {\tt jodw02@kth.se} and {\tt langmann@kth.se} \\
  [5mm]
\end{center}

\begin{abstract}
  We compute mean field phase diagrams of two closely related
  interacting fermion models in two spatial dimensions (2D).  The
  first is the so-called 2D \ttpV model describing spinless fermions
  on a square lattice with local hopping and density-density
  interactions. The second is the so-called 2D Luttinger model that
  provides an effective description of the 2D \ttpV model and in which
  parts of the fermion degrees of freedom are treated exactly by
  bosonization. In mean field theory, both models have a
  charge-density-wave (CDW) instability making them gapped at
  half-filling. The 2D \ttpV model has a significant parameter regime
  away from half-filling where neither the CDW nor the normal state
  are thermodynamically stable. We show that the 2D Luttinger model
  allows to obtain more detailed information about this mixed
  region. In particular, we find in the 2D Luttinger model a partially
  gapped phase that, as we argue, can be described by an exactly
  solvable model.
\end{abstract}

\bigskip
\section{Introduction} 
\label{sec1}
This is the second in a series of papers that aim to develop a method
to do reliable computations in a spinless 2D lattice fermion model of
Hubbard type; see \cite{EL0} for a concise summary. In the first paper
in this series \cite{EL1}, an effective model for the low-energy
physics of the lattice system was derived. It was shown that parts of
the fermion degrees of freedom in the model can be treated exactly
using bosonization.  In this paper, we apply mean field methods to the
remaining degrees of freedom. We present analytical and numerical
results showing that our method is useful for obtaining quantitative
physical information about the lattice fermions.

\subsection{Motivation} 
\label{sec1.1} 
The difficulty to do reliable computations in 2D lattice fermion
models of Hubbard-type has remained an outstanding challenge in
theoretical physics for many years. One can hope that a solution to
this problem will be a key step towards a satisfactory theory of
high-temperature superconductors \cite{Bonn}. One simple example is
the so-called 2D \ttpV model describing spinless fermions on a square
lattice with local hopping and density-density interactions; see
Section~\ref{sec2.1} for a precise definition.  One of us (EL)
proposed a particular partial continuum limit of this lattice system
\cite{EL0,EL1}. This leads to an interacting model of so-called nodal
fermions, which have linear band relations, coupled to so-called
antinodal fermions with hyperbolic band relations.  It was found that
this is a natural 2D analogue of the Luttinger model, not only in that
it arises as a continuum limit of the 2D \ttpV model, but also since
the nodal fermions can be bosonized and thus treated exactly. In
particular, it is possible to integrate out the bosonized nodal
fermions and thus obtain an effective model for the antinodal
fermions; see Section~\ref{sec2.2}.

In this paper we use mean field theory to address the question if and
when the antinodal fermions in this 2D Luttinger model have a
gap. These are key questions since if the antinodal fermions are
gapped they do not contribute to the low-energy physics. We then
obtain an effective Hamiltonian of nodal fermions that is exactly
solvable. We find a significant parameter regime away from
half-filling where this is indeed the case. As a motivation for our
work, we also present mean fields results for the 2D \ttpV model that
show that there is an interesting region away from half filling where
a direct application of mean field theory fails. We find that this
regime becomes accessible to mean field theory by using the 2D
Luttinger model.

We recall the key parameters of these two models.  The 2D \ttpV model
is characterized by the nearest-neighbor (nn) hopping constant $t>0$,
the next-nn (nnn) hopping constant $-t/2<t'<t/2$, the nn
density-density coupling strength $V/2>0$, and the filling factor
$0\leq \nu\leq 1$ (see Section~\ref{sec2.1} for more details). The 2D
Luttinger model depends on two additional parameters, $\kappa$ and
$Q$, which have the following significance. To derive the model, it is
assumed that there is an underlying Fermi surface that, when the
system is near half-filling, is a line segment in each nodal region.
These are the so called nodal Fermi surface arcs, and the parameter
$0\leq \kappa\leq 1$ determines the size of these arcs.  Furthermore,
the parameter $Q\approx \pi/2$ fixes the point about which the nodal
band relations are linearized; the details are given in
Section~\ref{sec2.2}. One important question addressed in this paper
is how to fix the parameter $Q$.

\subsection{Mean field phase diagrams} 
\label{sec1.2} 
Two-dimensional lattice fermion systems with repulsive interactions
are often insulators at half-filling, but away from half-filling there
are competing tendencies that lead to rich Hartree-Fock (HF) phase
diagrams. A well-known example is the 2D Hubbard model which, at
half-filling, has an insulating antiferromagnetic (AF) HF ground state
\cite{BLS}.  However, away from half-filling, unrestricted HF theory
yields intricate solutions that include domain walls, vortices,
polarons etc.\ imposed on an AF background; see e.g.\ \cite{unrestrHF}
and references therein. These solutions suggest that the pure AF state
is only stable at half-filling and that additional holes or particles
tend to distribute so as to perturb the AF state as little as
possible.  Furthermore, away from half-filling, these solutions break
translational invariance in a complicated manner and are highly
degenerate. One thus expects that the low-energy properties of the
lattice model away from half-filling should be dominated by
fluctuations between these degenerate solutions. Unfortunately, it is
difficult to formulate a useful low-energy effective model for this
situation.

Unrestricted HF theory is computationally demanding and thus
applicable only for moderate lattice sizes. However, it is possible to
find ``mixed'' regions in the phase diagram, i.e.\ regions with
intricate and highly degenerate HF solutions, already in mean field
theory. By the latter we mean HF theory restricted to states invariant
under translations by two sites \cite{LW0,LW}. This requires little
computational effort and thus allows to explore the full phase diagram
for arbitrarily large lattice sizes. In this manner one can identify
the regions in the phase diagram in which an exotic physical behavior
can be expected. It is remarkable how rich the resulting mean field
diagrams for the 2D Hubbard model are \cite{LW}.

\begin{figure}[!ht]
   \vspace{0.5in}
\begin{center}
\includegraphics[width=0.8\textwidth]{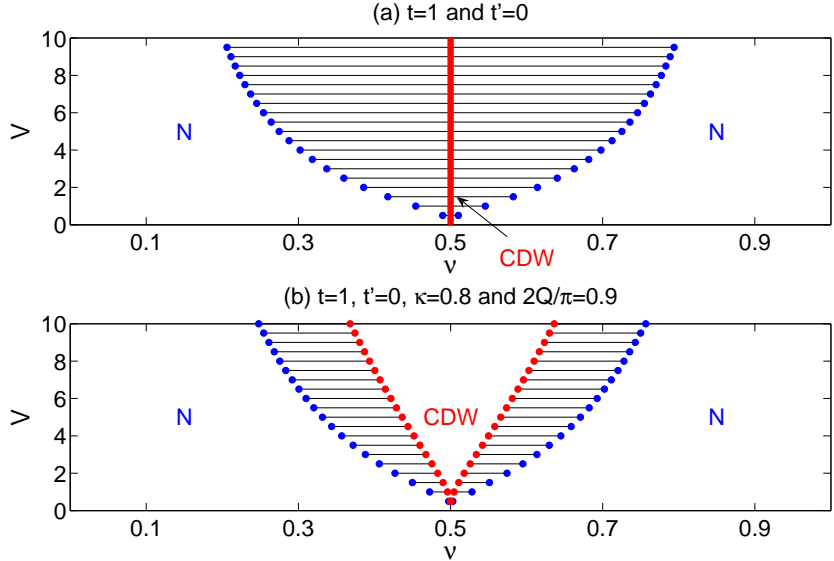} 
\end{center}
\caption{Comparison between the coupling ($V$) vs.\ filling ($\nu$)
  mean field phase diagrams of the 2D \ttpV model (a) and the 2D
  Luttinger model (b) for $t=1$ and $t'=0$. Shown are the
  charge-density-wave (CDW) and normal (N) phases as a function of
  coupling and filling at zero temperature. The regimes marked by
  horizontal lines are mixed, i.e.\ neither the CDW nor the N phase is
  thermodynamically stable. The CDW phase in (a) exists only at
  half-filling $\nu=0.5$. The CDW phase in (b) corresponds to a
  partially gapped phase.}
\label{Fig1}
\end{figure}

In this paper we present mean field phase diagrams for the 2D \ttpV
model, which is a spinless variant of the Hubbard model and therefore
somewhat simpler. We find a stable charge-density-wave (CDW) mean
field ground state at half-filling $\nu=0.5$, a translation-invariant
normal (N) state far away from half-filling, and extended doping
regions with mixed phases in-between. We indicate the mixed phases in
our phase diagrams by horizontal lines to emphasize that mean field
theory fails here, i.e.\ it does not allow any specific conclusions to
be drawn; see Figure~\ref{Fig1}(a). Put differently, the mixed regions
are "unknown territories" (analogous to white spots on ancient maps)
for mean field theory, but the knowledge of their existence is
nevertheless important physical information.

Our main result is to show that the 2D Luttinger model provides a tool
to explore these "unknown territories" using mean field theory for the
antinodal fermions. To be more specific, we compute mean field phase
diagrams for the effective antinodal Hamiltonian defined in
Section~\ref{sec2.2}. We find that the antinodal fermions indeed have
a CDW gap in a significant region of the parameter space, as
conjectured in \cite{EL0,EL1}.  We refer to this also as a CDW phase,
but we emphasize that, in general, it corresponds to a
\textit{partially gapped phase} of the 2D Luttinger model. There are
also the nodal fermions that are gapless, and these fermions can dope
the system even if the antinodal fermions remain gapped and
half-filled. In this way a large part of the mixed regions of the
phase diagram for the 2D \ttpV model is filled in; see
Figure~\ref{Fig1}(b).  We also study how sensitive the occurrence of a
partially gapped phase is to variations in the model parameters,
especially $\kappa$ and $Q$.

\subsection{Related work}
\label{sec1.3}
The derivation of the 2D Luttinger model was inspired by important
work of Mattis \cite{Mattis}, Schulz \cite{Schulz}, Luther
\cite{Luther2}, and Furukawa \textit{et al.} \cite{FRS}; see
\cite{EL1} for discussion and further references.

The phase diagram of the 2D \ttpV model, at and away from
half-filling, has been studied using various
techniques. Recent work close to ours is \cite{CRT} in which the
possibility of phase separation is investigated; see also
\cite{UV,KKK}. In particular, Figure~1(a) in \cite{CRT} is the same as
our Figure~\ref{Fig1}(a).  Note though, unlike \cite{CRT}, we do not necessarily
interpret the horizontally lined region in our Figure~\ref{Fig1}(a) as
a phase-separated state. Instead, we only conclude that the considered
mean field theory fails to give a stable homogeneous phase there.

One motivation for our work are experimental results on high
temperature superconductors. It is known from angle-resolved
photoemission spectroscopy \cite{ARPES} that these materials can have
an electronic phase in which parts of the underlying Fermi surface do
not have gapless excitations.  In particular, for hole-doped
materials, one finds near half-filling that the degrees of freedom in
the antinodal regions are gapped, while in each nodal region there is
an ungapped Fermi surface arc; see e.g.\ Figure~5 in \cite{A1} or
Figure~1 in \cite{A2}.\footnote{The experimental situation for
  electron-doped materials is different and its relation to the
  present work will not be discussed here.} We believe that this
suppression of accessible electronic states at the Fermi level seen in
experiments provide strong support for our approach. However, we
stress that the emphasis of the present work is mathematical, and a
confrontation of our results with experiments is postponed to future
work.

\subsection{Notation and conventions} 
\label{sec1.5} 
We use the symbol ``$\define$'' to emphasize that an equation is a
definition. We denote by $\Re{(c)}$ and $\Im{(c)}$ the real- and
imaginary parts of a complex number $c$, and $\overline{c}$ is its
complex conjugate. We write ``$a_{0,1}=A\pm B$'' short for ``$a_0=A+B$
and $a_1=A-B$'' etc.  We use bold symbols for matrices, e.g.\ $\id$ is
the identify matrix.  By ``$X=2.31(1)$'' we mean a numerical result
``$X=2.31\pm0.01$''.

The fermion models considered in this paper are defined by
Hamiltonians of the following generic type
\begin{equation}
  \label{Hgen} 
  H  =\sum_{kl} (t\pdag_{kl} -\mu\delta\pdag_{kl}) 
  c^\dag_k c\pdag_l + \sum_{klmn}v\pdag_{klmn}
c^\dag_kc^\dag_l c\pdag_n c\pdag_m
\end{equation} 
with fermion operators $c^{(\dag)}_k$ labeled by a finite number $\cN$
of one-particle quantum numbers $k$. Note that $\cN$ equals the number
of one-particle degrees of freedom that are included in the model.
Our normalization is such that
\begin{equation}
\label{normalization} 
\{ c\pdag_j,c^\dag_k\}=\delta_{jk}. 
\end{equation}
The model parameters $t_{kl}=\overline{t_{lk}}$ and
$v_{klmn}=\overline{v_{mnkl}}$ correspond to the matrix elements of
the kinetic energy and two-body interaction potential, respectively,
and $\mu$ is the chemical potential. The model is defined on a
fermion Fock space with a vacuum $|0\rangle$ annihilated by all $c_k$.
Expectation values with respect to a given state of the model (both
zero and non-zero temperatures) are denoted by $\langle\cdot\rangle$.
The state can be either the exact thermal equilibrium state or an
approximate Hartree-Fock state; it will always be clear from the
context which is meant. The inverse temperature is denoted by
$\beta>0$.

\subsection{Plan of paper} 
\label{sec1.4}
The two models we consider are defined in Section~\ref{sec2}. Some
results presented here are (minor) generalizations of the
corresponding ones in \cite{EL1}, as further elaborated in
Appendix~\ref{appA}.  Section~\ref{sec3} explains the method we use to
compute mean field phase diagrams, with some technical details
deferred to appendices~\ref{appB} and \ref{appC}. Our results are
given and discussed in Section~\ref{sec4}.  Section~\ref{sec5} gives
some closing remarks.

\section{Models}
\label{sec2}
In this section, we define the two studied models and discuss the
relation between them.  We note in passing that the notation used here
is slightly different from that in \cite{EL1}; see
Appendix~\ref{appA1} for details.

\subsection{2D \ttpV model}
\label{sec2.1} 
\begin{figure}[ht!]
\begin{center}
\includegraphics[width=0.75\textwidth]{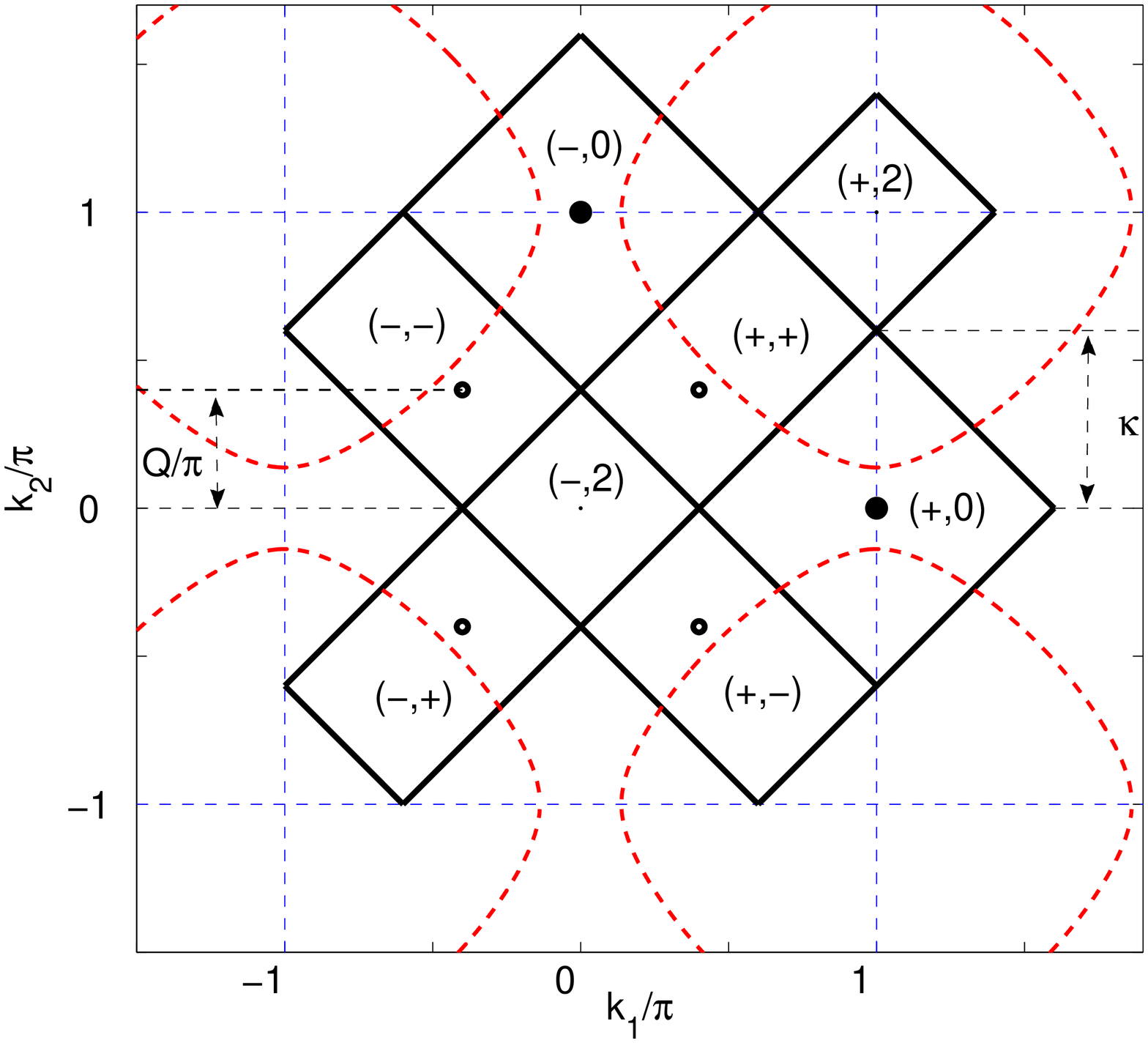}
\end{center}
\caption{Fermi surface for the tight-binding band relation in
  \Ref{eps0} with $t=1$, $t'=-0.2$ and $\mu = -0.51(1)$ (bent dashed
  curves). Also indicated are the eight regions labeled by $(r,s)$ and
  discussed in the main text (large squares for $r=\pm$, $s=0$,
  rectangles for $r=\pm$, $s=\pm$, and small squares for $r=\pm$,
  $s=2$).  Note that the Brillouin zone is equal to the union of these
  regions. The two antinodal points $\vQ_{r,0}$ and four nodal points
  $\vQ_{r,\pm}$ are depicted by dots and rings, respectively.  The
  parameters used in the plot are $\kappa=0.6$ and $Q/\pi=0.4$, and
  $\mu$ corresponds to $\tQ /\pi = 0.45(1)$.}
\label{Fig2}
\end{figure}

The 2D \ttpV model describes spinless fermions on a square lattice
with $L^2$ sites and lattice constant $a=1$. The fermions hop with
amplitudes $t$ and $t'$ between nn and nnn sites, and fermions on nn
sites interact with a density-density interaction of strength $V/2$.
The Hamiltonian is given in Fourier space by
\begin{equation}
\label{H}
\HttpV = \sum_{\vk\in\BZ}[\eps(\vk)-\mu]c^\dag(\vk)c(\vk) + \frac{V}{2L^2}
\sum_{\vk_j\in\BZ} v_{\vk_1,\vk_2,\vk_3,\vk_4} 
c^\dag(\vk_1)c^\dag(\vk_3)c(\vk_4) c(\vk_2)
\end{equation} 
with the tight-binding band relation ($\vk = (k_1,k_2)$)
\begin{equation}
\label{eps0} 
\eps(\vk)=-2t[\cos(k_1)+\cos(k_2)] - 4t'\cos(k_1)\cos(k_2) 
\end{equation}
and the interaction vertex\footnote{The sum over $\vn$ in \Ref{v}
  takes into account possible umklapp processes.}
\begin{equation}
\label{v} 
v_{\vk_1,\vk_2,\vk_3,\vk_4} = u(\vk_1-\vk_2)\sum_{\vn\in\Z^2} 
\delta_{\vk_1-\vk_2+\vk_3-\vk_4,2\pi\vn}
\end{equation} 
with
\begin{equation}
\label{u}
u(\vp)= \cos(p_1)+\cos(p_2). 
\end{equation} 
The fermion operators $c^{(\dag)}(\vk)$ are labeled by momenta $\vk$
in the Brillouin zone
\begin{equation}
\label{BZ}
\BZ \define \bigl\{ \vk=(k_1,k_2)\; \bigl| \; 
-\pi < k_{j} <\pi\; \mbox{, }\; 
k_{j} =\mbox{$\frac{2\pi}L$}(n_{j}+\half)
\;\mbox{, }\; 
n_{j}\in\Z\; \mbox{, }\; j=1,2 \bigr\}.  
\end{equation} 
Other conventions used are explained in Section~\ref{sec1.5} (with the
$k$ there corresponding to our $\vk$ here).  Note that the number of
different one-particle quantum numbers $\vk$ equals the system volume,
$\cN=L^2$.  The chemical potential $\mu$ is to be determined such that
the fermion density
\begin{equation}
\label{nu} 
\nu\define \frac1{L^2}\langle N\rangle,\quad 
N \define \sum_{\vk\in\BZ} c^\dag(\vk)c(\vk)
\end{equation} 
has a fixed specified value.  We refer to $\nu$ as \textit{filling
  factor} or filling, and $\nu-0.5$ is called \textit{doping}. The
filling lies in the range $0\leq \nu\leq 1$, and $\nu=0.5$ corresponds
to half-filling. We always assume $-t/2<t'<t/2$, $V>0$, and that $L/2$
is an integer. To simplify notation we set $t=1$ in all figures, i.e.\
energies are measured in units of $t$.  Figure~\ref{Fig2} shows the
Brillouin zone of this model and a typical example of a
non-interacting Fermi surface defined by $\eps(\vk)=\mu$.

Note that we use anti-periodic boundary conditions and a
large-distance cutoff different to that used when deriving the 2D
Luttinger model in \cite{EL1}.  This is legitimate since we are
interested in the thermodynamic limit $L\to\infty$, and finite size
effects are negligible for the system sizes $(L\geq 100)$ we use in
our numerical computations.

The invariance of the 2D \ttpV model under particle-hole
transformations provides an important guide for us.  The interested
reader can find more details in Appendix~\ref{appA1}.

\subsection{2D Luttinger model}
\label{sec2.2} 
A detailed derivation of the 2D Luttinger model and its partial
solution by bosonization were given in \cite{EL1}. Here we first
describe this model and then summarize the results from \cite{EL1}
that we actually need.

The 2D Luttinger model involves six fermion flavors labeled by a pair
of indices $(r,s)$ with $r=\pm$ and $s=0,\pm$.  These flavors
correspond to different regions in the Brillouin zone of the 2D \ttpV
model, as shown in Figure~\ref{Fig2} (two large squares for $s=0$ and
four large rectangles for $s=\pm$).  The sizes of these regions are
determined by a parameter $\kappa$ in the range $0\leq \kappa\leq 1$
(we used $\kappa=0.6$ in Figure~\ref{Fig2}). There are two more
fermion flavors with $r=\pm$ and $s=2$ (two small squares in
Figure~\ref{Fig2}), but in this paper we assume that the parameters
are such that the $s=2$ fermions are far away in energy from the Fermi
level and thus their dynamics can be ignored \cite{EL1} (we discuss
this point in Section~\ref{sec5}, Remark~2).

The momenta in the regions with $s=0$ can be written as
$\vQ_{\pm,0}+\vk$ with
\begin{equation}
\label{vQr0}
\vQ_{+,0}=(\pi,0),\quad \vQ_{-,0}=(0,\pi)
\end{equation}
and $\vk$ in
\begin{equation}
\label{BZa}
\Lambda^*_a \define \bigl\{\vk\in\BZ\;\bigl|\;  
|k_1\pm k_2|\leq \kappa\pi \; \mbox{, } \; 
k_1\pm k_2 = \mbox{$\frac{2\sqrt{2}\pi}{L}$}(n_\pm +\half) \; 
\mbox{, }\; n_\pm\in\Z  \bigr\}.   
\end{equation}
Close to the points $\vQ_{r,0}$, $r=\pm$, the band relation in
\Ref{eps0} can be well approximated by the lowest-order non-trivial
terms in a Taylor series expansion $\eps(\vQ_{r,0}+\vk)\approx 4t' +
\EE_{r}(\vk)$ with
\begin{equation}
\label{Epm} 
\EE_{r}(\vk) = rt(k_1^2-k_2^2) - 2t'(k_1^2+k_2^2), 
\end{equation} 
i.e.\ $\vQ_{r,0}$ correspond to saddle points of $\eps(\vk)$ if
$-t/2<t'<t/2$. This also explains why we impose these bounds on $t'$.
Similarly, the band relation for fermion degrees of the freedom
corresponding to $s=\pm$ can be well approximated by bands linear in
$(k_1 \pm k_2)$ close to the points $\vQ_{r,\pm}=(rQ,\pm rQ)$ with $Q$
another free parameter ($r=+$ or $-$). We use standard terminology
\cite{ARPES} and refer to the fermions with $s=0$ as
\textit{antinodal} and the fermions with $s=\pm$ as \textit{nodal},
respectively.  In Figure~\ref{Fig2} the antinodal points $\vQ_{r,0}$
are indicated by dots and the nodal points $\vQ_{r,\pm}$ by circles.

The 2D Luttinger model is defined by a Hamiltonian of the form
$H_n+H_a+H_{na}$ where $H_n$ and $H_a$ include terms depending only on
the nodal and antinodal fermions, respectively, and $H_{na}$ are
interaction terms with both kinds of fermions. It is obtained from the
2D \ttpV model by certain approximations that amount to a particular
partial continuum limit \cite{EL1}.  A key assumption is that there is
an underlying Fermi surface in the nodal regions $s=\pm$ consisting of
line segments (``Fermi surface arcs'') and containing the nodal points
$(r\tQ ,\pm r\tQ )$ for some $\tQ \approx \pi/2$ determined by $\mu$,
$\kappa$ and $Q$; see \Ref{fixmu}. In \cite{EL1} we fixed $\mu$ by the
condition $\tQ=Q$, but in the present paper we work in the grand
canonical ensemble and thus allow $\mu$ (i.e. $\tQ$) to be arbitrary
in intermediate steps of our computations. No assumption is made on
the Fermi surface in the antinodal regions.

As explained in \cite{EL1}, Sections~2 and 6.2, we need to restrict
ourselves to parameters such that
\begin{equation}
\label{restrict} 
  (1-\kappa)\pi/2<Q<(1+\kappa)\pi/2\; \mbox{ and }\;  
  0\leq  \frac{V(1-\kappa)\sin(Q)}{2\pi [t+2t'\cos(Q)]}<1 
\end{equation} 
with the same bound for $\tQ$ as for $Q$. Moreover, we require $\tQ$
and $Q$ to be different from $\pi/2$, since otherwise one has
additional back-scattering terms in the 2D Luttinger model which spoil
a simple treatment of the nodal fermions using bosonization; see
\cite{EL1}. These conditions define the parameter regime of interest
to us.

The nodal fermions in the 2D Luttinger model can be bosonized and
integrated out exactly. This yields an effective model for the
antinodal fermions $c_\pm^{(\dag)}(\vk)\define
c^{(\dag)}_{\phantom\pm}(\vQ_{\pm,0}+\vk)$ that, in the local-time
approximation \cite{EL1}, is given by a Hamiltonian of the form
\begin{equation}
\label{HLutt} 
\HLutt = \sum_{\vk\in\Lambda^*_a}\sum_{r=\pm}[ \EE_{r}(\vk)-\mu_a] 
c^\dag_r(\vk) c\pdag_r(\vk) + \frac{2 g_a}{L^2} 
\sum_{\vk_j\in\Lambda^*_a} \delta_{\vk_1-\vk_2+\vk_3-\vk_4,\vzero}\, 
c^\dag_+(\vk_1) c^\dag_-(\vk_3) c\pdag_-(\vk_4)
c\pdag_+(\vk_2).
\end{equation}
This Hamiltonian is also of the generic type discussed in
Section~\ref{sec1.5} (with the one-particle quantum numbers $k$ used
there to be identified with $(r,\vk)$ here). Note that there are
$(\kappa L)^2/2$ different momenta $\vk$, and that the number of
one-particle degrees of freedom in the model is $\cN=(\kappa
L)^2$. The filling factor of the antinodal fermions is therefore
\begin{equation}
\label{nua} 
\nu_a \define   \frac1{(\kappa L)^2} \langle N_a\rangle,\quad    N_a \define  
\sum_{\vk\in\Lambda^*_a}\sum_{r=\pm}
c^\dag_r(\vk) c\pdag_r(\vk), 
\end{equation}
while the total filling factor of the 2D Luttinger model is (including
the nodal fermions etc.)
\begin{equation}
\label{nu1} 
  \nu = \half + (1-\kappa)(\tQQPi-1) + \kappa^2(\nu_a - \half ) ; 
\end{equation}
see \cite{EL1}. It is $\nu$ in \Ref{nu1} that is to be identified with
the filling factor of the 2D \ttpV model.

A main result in \cite{EL1} are explicit formulas for the parameters
$\mu_a$ and $g_a$ in terms of the other model parameters:
\begin{eqnarray} 
\label{mua} 
\mu_a = \mu -2V\!\nu -4t' + g_a\nu_a\kappa^2
\end{eqnarray} 
and 
\begin{equation}
\label{ga} 
g_a = 2V-\geff
\end{equation}
with 
\begin{equation}
\label{geff} 
\geff = \frac{V^2(1-\kappa)}{\sin(Q)\pi[t + 2t'\cos(Q) +
  \frac{V}{\pi}(1-\kappa)\sin(Q)]} .  
\end{equation} 
The parameter $\tQ$ can be computed from the following identity
\begin{equation}
\label{fixtQ} 
\mu = \sqrt2 v_F(\tQ-Q) + 2V\!\nu -4t\cos(Q)-4t'\cos^2(Q) -2VC\cos(Q) 
\end{equation}
with
\begin{equation}
\label{v_F}
v_F=2\sqrt2\sin(Q)[t+2t'\cos(Q)]
\end{equation} 
and where we have introduced a convenient short-hand notation
\begin{equation}
\label{C}
C \define (1-\kappa)\cos(Q)(\tQQPi-1) +\half(1-\kappa)^2. 
\end{equation}

The constant $-\geff$ corresponds to a renormalization of the bare
antinodal interaction $2V$ and arises from integrating out the
bosonized nodal fermions. 

The 2D \ttpV Hamiltonian is equivalent (in a low-energy approximation)
to the effective antinodal Hamiltonian in \Ref{HLutt} only if one
takes into account the additive constant (see Appendix~\ref{appA2})
\begin{equation}
\label{cEa}
\cE_a = \cE_{kin}+\cE_\uno+\cE_{int}+\cE_n  
\end{equation}
with
\begin{equation}
\begin{split}
\label{cEkin}
  \cE_{kin}/L^2 = \half(1-\kappa)^2 \bigl(
  -4(t+t')+\third(t+2t')(1-\kappa)^2\pi^2
  \bigr)+ (1-\kappa)(\tQQPi-1+\kappa)\\
   \times\bigl( \sqrt 2 v_F \left( {\tQ/2 - Q + \pi\left( {1 - \kappa } 
   \right)/4} \right) - 4t\cos(Q) - 4t'\cos ^2(Q) \bigr),
\end{split} 
\end{equation} 
\begin{equation}
\label{cE1} 
\cE_\uno/L^2 = 
(\mu-2V\!\nu)\nu^{\phantom 2}_a\kappa^2 + \half g_a\nu_a^2\kappa^4, 
\end{equation} 
and 
\begin{equation}
\label{cEint}
\cE_{int}/L^2 = -\mu\nu + V \nu^2+ V[\kappa(1-\kappa)\cos(Q)]^2 -VC^2. 
\end{equation}
The constant $\cE_n$ is derived in \cite{EL1}, but in the present
paper we only need that it is independent of the chemical potential:
\begin{equation}
\label{cEn}
\frac{\partial(\cE_n/L^2)}{\partial\mu} = 0 .  
\end{equation}

As we show later, the parameter $Q$ can be fixed by the
self-consistency condition $Q=\tQ$. Thus the effective antinodal model
contains one more free parameter as compared to the 2D \ttpV model,
namely $\kappa$.

As already mentioned, the results in \cite{EL1} are restricted to the
special case in which $\mu$ is explicitly fixed by the condition
$\tQ=Q$, and they are written in a slightly different form.  The
interested reader can find details about how to obtain the results
given here from the ones in \cite{EL1} in Appendix~\ref{appA2}.

It is not essential to work with the Taylor expansion of the antinodal
band relations, and one can equally well use the full band relations
\begin{equation}
\label{Epm0}
\EE_r^{(0)}(\vk) = \eps(\vQ_{r,0}+\vk) - \eps(\vQ_{r,0})
\end{equation} 
instead of \Ref{Epm}.  As we will discuss, the results for the band
relations in \Ref{Epm} and \Ref{Epm0} agree quite well for smaller
values of $\kappa$, but for $\kappa$ close to one there are some
quantitative differences.  Furthermore, in the derivation of the 2D
Luttinger model, certain approximations were done on the interaction
vertex of the 2D \ttpV model; see Section~5 in \cite{EL1}. It was
argued that this had no important consequences for low-energy
scattering processes. We note here that these approximations were only
necessary for processes that include nodal fermions, and it would be
possible to use the full interaction vertex for processes involving
just antinodal fermions. With this, and using the full band relation
\Ref{Epm0}, one could derive a refined 2D Luttinger model for which 
the 2D \ttpV model is recovered by setting $\kappa=1$.

We finally mention that the local-time approximation in \cite{EL1} was
only done for simplicity, and it is possible to generalize our
treatment here to take into account the full time dependent
interaction. We hope to come back to this in the future.

\section{Method}
\label{sec3}
In this section, we discuss the mean field Hamiltonians used for
deriving phase diagrams and the procedure that allows us to identify
the mixed regions. More explicit details can be found in
appendices~\ref{appB} and \ref{appC}.

The fermion models considered in the previous section are defined by
Hamiltonians of the type \Ref{Hgen}.  Conventional HF theory at zero
temperature and fixed particle number $N_0$ for such models amounts to
considering the set of all variational states of the form $\eta =
c^\dag(f_1)c^\dag(f_2)\cdots c^\dag(f_{N_0})|0\rangle$ with
$c^\dag_{\phantom i}(f_j)=\sum_{k} (f_j)_k\pdag c^\dag_k$ and $f_j$
orthonormal one-particle states.\footnote{To be more precise:
  the $f_j$ are vectors in $\mathbb{C}^{\cN}$ with components
  $(f_j)_k$, and $\sum_k \overline{( f_i)_k} (f_j)_k=\delta_{ij}$.}
Using standard terminology, we refer to the states $\eta$ as Slater
determinants.  The $f_j$ are to be chosen so as to minimize the HF
energy
\begin{equation}
\label{EHF} 
\OmHF \define  \langle \eta, H \eta \rangle . 
\end{equation} 
It is straightforward to compute an explicit formula of $\OmHF$ in
terms of the one-particle density matrix
$\gamma_{ij}=\sum_{k=1}^{N_0}(f_k)_i\overline{(f_k)_j}$; see
Appendix~\ref{appB}.

In the present paper, we work at (mainly) small but non-zero
temperatures (unless otherwise indicated $\beta=10^5$).  This means
that we minimize the full grand canonical potential (including the
entropy) with respect to all HF Gibbs states; see Appendix~\ref{appB}
for a full discussion. We will however set the temperature to zero in
the current section to not burden the presentation.

\subsection{2D \ttpV model}
\label{sec3.1}
It is convenient to choose the Slater determinant $\eta$ as ground
state of a reference Hamiltonian
\begin{equation}
\label{HR}
H_{\HFG} = H_0 + \sum_{\vk,\vk'} w(\vk,\vk') c^\dag(\vk) c(\vk')  
\define  \sum_{\vk,\vk'} h_w(\vk,\vk') c^\dag(\vk) c(\vk')  
\end{equation}
with $H_0$ the non-interacting Hamiltonian obtained from the one in
\Ref{H} by setting $V=0$, and $w(\vk,\vk')=\overline{w(\vk',\vk)}$ the
matrix elements of the HF potential $w$. This allows to parametrize HF
theory by the one-particle Hamiltonian $h_w$ defined in \Ref{HR}. As
explained below, it is important to use the grand canonical ensemble,
i.e.\ to fix the particle number by adjusting the chemical potential
$\mu$ \cite{LW}. One thus obtains
\begin{equation}
\label{one-pdm-zeroT}
\gamma(\vk,\vk') = \sum_k \theta(-e_k)f_k(\vk) \overline{f_k(\vk')} 
\end{equation} 
with $e_k$ and $f_k$ the eigenvalues and corresponding orthonormal
eigenvectors of $h_w$ and $\theta$ the Heaviside function.  Note that
the chemical potential is included in $e_k$.

Unrestricted HF theory amounts to determining the HF potential $w$
that minimizes $\OmHF$ under the filling constraint
\begin{equation} 
\label{nuHF} 
\nu = -\frac{\partial(\OmHF/L^2)}{\partial\mu} \, . 
\end{equation} 
This method is computationally demanding and thus restricted to small
system sizes.

By \textit{mean field theory} we mean the restriction of HF theory to
states that are invariant under translations by two sites. As
explained in Appendix~\ref{appC1}, this corresponds to considering the
following restricted set of HF potentials:
\begin{equation}
\label{MF} 
w(\vk,\vk') = \bigl(q_0 + q_1 [\cos(k_1)+\cos(k_2)]\bigr)\,  
\delta_{\vk,\vk'}+ \Delta \, \delta_{\vk,\vk'+\vQ} 
\end{equation} 
with $\vQ=(\pi,\pi)$ and three real variational parameters $q_0$,
$q_1$ and $\Delta$. For $\Delta=0$ this corresponds to a
normal (N) state that is translation invariant, and for
$\Delta \neq 0$ one has a charge-density-wave (CDW) state for
which translation invariance is broken down to translations by two
lattice sites.  The computational problem is now very easy: there are
only three variational parameters, and $\OmHF$ can be computed
analytically by Fourier transformation; the interested reader can find
details in Appendix~\ref{appC1}.

\begin{figure}[ht!]
\begin{center}
\includegraphics[width=0.75\textwidth]{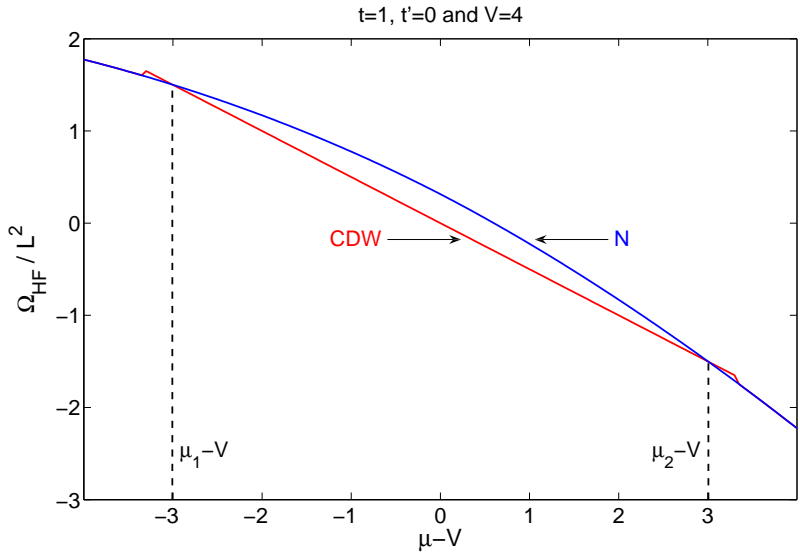}
\end{center}
\caption{Hartree-Fock energies of the charge-density-wave (CDW) and
  normal (N) states vs.\ chemical potential $\mu$ for the 2D \ttpV
  model at $t=1$, $t'=0$, $V=4$. The energy $\OmHF$ of the
  thermodynamically stable phase is given by the minimum of these
  curves. The constant slope of the CDW curve in $\mu_1<\mu<\mu_2$
  implies that the CDW phase can exist only at half-filling, and the
  kinks at $\mu=\mu_{1,2}$, where the system changes from the CDW to
  the N phase, imply mixed regions where neither the CDW nor the N
  phase is thermodynamically stable. (A convenient overall constant
  has been added to $\OmHF$.)}
\label{Fig3}
\end{figure}

It is important to note that, by working with the grand canonical
ensemble, we not only can detect variational ground states given by
Slater determinants $\eta_{\CDW}$ (with $\Delta \neq 0$) or
$\eta_{\N}$ (with $\Delta =0$), but it is also possible that the state
with lowest energy is mixed and of the form
\begin{equation}
\label{mixed}
\lambda|\eta_{\CDW}\rangle\langle\eta_{\CDW}| 
+ (1-\lambda)|\eta_{\N}\rangle\langle\eta_{\N}| 
\end{equation} 
for some $0<\lambda<1$.\footnote{We identify a Slater determinant
  $\eta$ with the state $|\eta\rangle\langle\eta|$.} In general, the
two Slater determinants correspond to different filling factors
$\nu_{\CDW}$ and $\nu_{\N}$, and the state in \Ref{mixed} corresponds
to total filling
\begin{equation} 
\label{mixednu} 
\nu=\lambda\nu_{\CDW} + (1-\lambda)\nu_{\N}.
\end{equation} 
The possibility of obtaining \Ref{mixed} can be seen by computing the
HF energies for the CDW and N states, $\OmHF^X$ for $X=\CDW$ and $\N$,
as functions of the chemical potential $\mu$; see e.g.\
Figure~\ref{Fig3} for $t=1$, $t'=0$ and $V=4$. One finds two different
regimes: there is an interval $\mu_1<\mu<\mu_2$ where
$\OmHF^{\CDW}<\OmHF^{\N}$, i.e.\ the system clearly has a CDW mean
field ground state $\eta_{\CDW}$; for $V/t=4$ and $t'=0$ we find
$\mu_2-V=V-\mu_1=3.01(1)$.\footnote{The accuracy of $\mu_2$ reached in
  our computations is actually greater than what the numerical error
  given here suggests. A similar remark applies to other numerical
  results given in this paper.} The second kind of regime is for
$\mu<\mu_1$ and $\mu>\mu_2$ where one finds $\OmHF^{\CDW}<\OmHF^{\N}$,
i.e.\ a N ground state $\eta_{\N}$. However, not all doping values can
be realized in this way: according to \Ref{nuHF} we can obtain the
doping $\nu$ from the slope of
$$
\OmHF^{\phantom X}=\min(\OmHF^{\CDW}, \OmHF^{\N}),
$$
but this is not continuous at $\mu=\mu_{j}$, $j=1,2$, where the CDW
state changes to the N state.  The mean field phases can be determined
by the following four doping values,
\begin{equation}
  \nu_{X,j}\define  \left. 
    -\frac{\partial (\OmHF^X/L^2)}{\partial\mu}\right|_{\mu=\mu_j} ,\quad
  j=1,2,\quad X=\CDW, \N
\end{equation} 
as follows. The system is in the CDW phase in the doping regime
$\nu_{\CDW,1}\leq\nu\leq\nu_{\CDW,2}$ and in the N phase for
$\nu\leq\nu_{\N,1}$ and $\nu\geq \nu_{\N,2}$. However, in the regions
$\nu_{\N,1}<\nu<\nu_{\CDW,1}$ and $\nu_{\CDW,2}<\nu<\nu_{\N,2}$ the
mixed state in \Ref{mixed}, with $\lambda$ determined by
\Ref{mixednu}, has lower energy than either of the pure Slater
states. We refer to the latter as a \textit{mixed phase}. We emphasize
that mixed phases occur in large parts of the phase diagram; e.g.\ for
$t=1$, $t'=0$ and $V=4$ we find $\nu_{\CDW,1}=\nu_{\CDW,2}=0.5$ and
$\nu_{\N,1}=1-\nu_{\N,2} = 0.30(1)$.

The results described above have the following physical interpretation
(we only discuss the regions close to $\mu=\mu_2$ since the other one
is similar). In the CDW phase one has the effective band relations
(see Appendix~\ref{appC1})
\begin{equation}
\label{CDWbands}
e_\pm(\vk) = -4t'\cos(k_1)\cos(k_2) + q_0 -\mu \pm 
\sqrt{(-2t+q_1)^2[\cos(k_1)+\cos(k_2)]^2 + \Delta^2 }.
\end{equation} 
This shows that the CDW phase has a band gap $2|\Delta|$, and as long
as $\mu$ is in this gap, changing it cannot affect doping. Thus the HF
energy is a linear function of $\mu$ with slope $\nu_{\CDW}=0.5$
(half-filling) in this region. The N phase is not gapped and doping
can be monotonically increased by increasing $\mu$, and therefore the
HF energy is a strictly concave function of $\mu$. Thus, as we try to
increase doping by increasing $\mu$ in the CDW phase, the HF energy of
the N phase decreases faster than the HF energy of the CDW phase, and
when both energies become equal at $\mu=\mu_2$ the doping $\nu_{\N,2}$
in the N phase is significantly larger than the doping $\nu_{\CDW,2}$
in the CDW phase.

A possible interpretation of the mixed state in \Ref{mixed} is a
phase-separated state in which parts of the system are in the CDW
phase and parts in the N phase \cite{LW,CRT}.  We can therefore
conclude that, for $\nu_{\N,2}<\nu<\nu_{\CDW,2}$, a phase-separated
state has lower variational energy than any simple mean field
state. However, we emphasize that the occurrence of a mixed phase does
not necessarily imply phase separation, but it nevertheless proves
that a true HF ground state is very different from any state that can
be described by a simple mean field ansatz \Ref{MF} (a true HF ground
state can in principle be found by unrestricted HF theory).  Thus mean
field theory allows to determine those regions in phase space where
non-conventional physics (not describable by mean field theory) is to
be expected.

It is interesting to note that, for non-zero $t'$, it is possible to
have a CDW phase also away from half-filling. This can be seen by
computing a plot similar to Figure~\ref{Fig3} but with $t'=-0.2$, for
example.  One still finds that the CDW energy $\OmHF^{\CDW}$ as a
function of $\mu$ is a straight line in most of the interval
$\mu_1<\mu<\mu_2$, but as $\mu$ approaches $\mu_2$ from the left, this
curve starts to bend so that $\nu_{\CDW,2} = 0.53(1)>0.5$. There is no
bending of $\OmHF^{\CDW}$ close to $\mu=\mu_1$, however, and
$\nu_{\CDW,1}=0.5$. Thus at $t=0$, $t'=-0.2$, $V=4$ it is possible to
dope the CDW state on the particle side (i.e.\ for $\nu>0.5$) but not
on the hole side ($\nu<0.5$). This shows that the parameter $t'$
affects the mean field phase diagrams both quantitatively and
qualitatively.

Our results for the 2D \ttpV model were obtained with MATLAB using the
system size $L=100$ (i.e. $100^2$ lattice sites). We checked that this
is large enough so that finite size effects are essentially
negligible. However, we note that some phase boundaries are slightly
affected by finite size effects even at this system size, as discussed
in more detail below.

\subsection{2D Luttinger model}
\label{sec3.2}
We use HF theory as explained for the 2D \ttpV model in the previous
section. The reference Hamiltonian can now be written as (using a
convenient matrix notation)
\begin{equation}
\label{HqfLUTT} 
H_{\HFG} = \sum_{\vk\in\Lambda^*_a}
 \bigl( c^\dag_+(\vk), c^\dag_-(\vk) \bigr)
\left(\!\!\begin{array}{cc}
      \EE_+(\vk)+q_0+q_1-\mu_a  & \Delta
      \\ \overline{\Delta} &
      \EE_-(\vk)+q_0-q_1-\mu_a    
\end{array}\!\!\right)
\left(\!\!\begin{array}{l} c_+(\vk)
    \\ c_-(\vk) \end{array}\!\!\right)
\end{equation} 
with variational parameters $q_0$, $q_1$ and $\Delta$.

The grand canonical potential corresponding to this reference
Hamiltonian, evaluated for the Hamiltonian in \Ref{HLutt}, is denoted
by $\Omega_a$; see Appendix~\ref{appC2.3} for explicit formulas. It is
important to note that this is only the antinodal contribution and
that the total grand canonical potential of the 2D Luttinger model is
\begin{equation}
\label{grandCanonicalLuttinger}
\Omega = \Omega_a+\cE_a
\end{equation}
with the energy constant $\cE_a$ in \Ref{cEa}--\Ref{cEn} taking into
account the contributions from the other fermion flavors $(r,s)$,
$r=\pm$ and $s=\pm,2$. Note that $\Omega_a$ is a function of $\mu_a$
(rather than $\mu$), and that the filling constraint following from
our general discussion of HF theory in Appendix~\ref{appB} is
\begin{equation}
\label{nuaHF} 
\nu_a = -\frac1{(\kappa L)^2}\frac{\partial\Omega_a}{\partial\mu_a}. 
\end{equation} 
On the other hand, the filling constraint in \Ref{nuHF} should still
hold true but with $\OmHF$ replaced by $\Omega$ in
\Ref{grandCanonicalLuttinger}. This implies the following consistency
condition
\begin{equation}
  \label{consistency}
\frac{\partial(\cE_a/L^2)}{\partial\mu} = -\nu +
\kappa^2\nu_a\frac{\partial\mu_a}{\partial\mu}  
\end{equation}
which must be fulfilled for the parameters $\mu_a$ and $\cE_a$ given
in Section~\ref{sec2.2}. This identity is non-trivial and provides an
important check of our computations. We therefore include details of
its verification in Appendix~\ref{appC2.4}.

\begin{figure}[!ht]
   \vspace{0.5in}
\begin{center}
\includegraphics[width=0.8\textwidth]{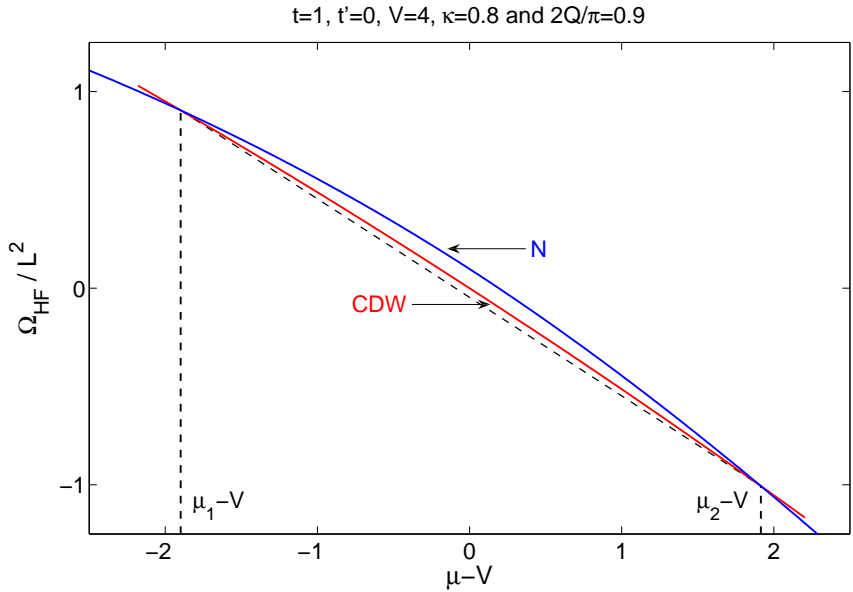}
\end{center}
\caption{Hartree-Fock energies vs.\ $\mu$ as in Figure~\ref{Fig3} but
  now for the 2D Luttinger model at $\kappa=0.8$ and $Q=0.45\pi$
  ($V=4$, $t=1$, $t'=0$). Shown are the energies of the antinodal
  charge-density-wave (CDW) and normal (N) phases. Again there is an
  interval $\mu_1<\mu<\mu_2$ for which the CDW energy is lower, but
  unlike Figure~\ref{Fig3}, this energy now deviates from a straight
  line (dashed in the figure).  This is due to the nodal fermions that
  change the total filling $\nu$ even though the antinodal fermions
  are half-filled and gapped. This kind of plot proves the existence
  of a partially gapped mean field phase in the 2D Luttinger model.}
\label{Fig4}
\end{figure}

The calculation of the grand canonical potential of the antinodal CDW
phase is done as follows (the normal phase is treated
identically). For fixed $\mu$, we make an ansatz for the antinodal
filling $\nu_a$. Solving \Ref{fixtQ} for $\tQ$ then gives us $\mu_a$
using \Ref{mua}. We proceed by minimizing the grand canonical
potential \Ref{grandCanonicalLuttinger} with respect to the
variational parameters $q_0$, $q_1$ and $\Delta$. This in turn gives
us a specific value for the antinodal filling to be compared with our
initial guess. We repeat the above procedure until a self-consistent
solution is obtained for $\nu_a$. An example of a resulting curve for
$\Omega$ vs. $\mu$ is given in Figure~\ref{Fig4}.

For the numerical calculation of $\Omega_a$, we fix the number of
momenta in the antinodal Fourier space regions \Ref{BZa} to
$6400$. Then $L=80\sqrt{2}/\kappa$, which is large enough so that
finite size effects are smaller than the symbol size in our
figures. Furthermore, at this system size we can safely replace the
Riemann sums in Appendix~\ref{appA2.3} by integrals.

\section{Phase diagrams} 
\label{sec4}
In this section, we present and discuss mean field phase diagrams for
the 2D \ttpV model and the 2D Luttinger model.

\subsection{2D \ttpV model} 
\label{sec4.1} 
\begin{figure}[!ht]
\begin{center}
\includegraphics[width=0.7\textwidth]{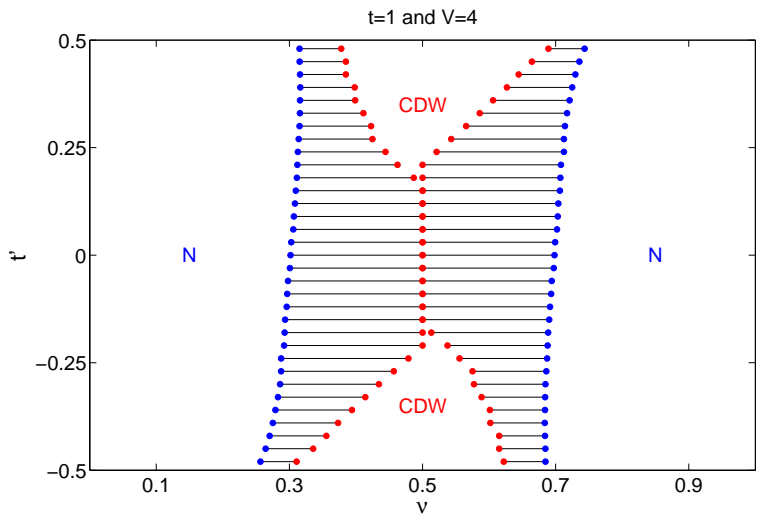}
\end{center}
\caption{Mean field phase diagram of the 2D \ttpV model at zero
  temperature: next-nearest-neighbor hopping $t'$ vs.\ filling $\nu$
  at $t=1$ and $V=4$. Shown are the charge-density-wave (CDW), normal
  (N), and mixed (horizontal lines) phases.}
\label{Fig5}
\end{figure}

\begin{figure}[!ht]
   \vspace{0.5in}
\begin{center}
\includegraphics[width=0.7\textwidth]{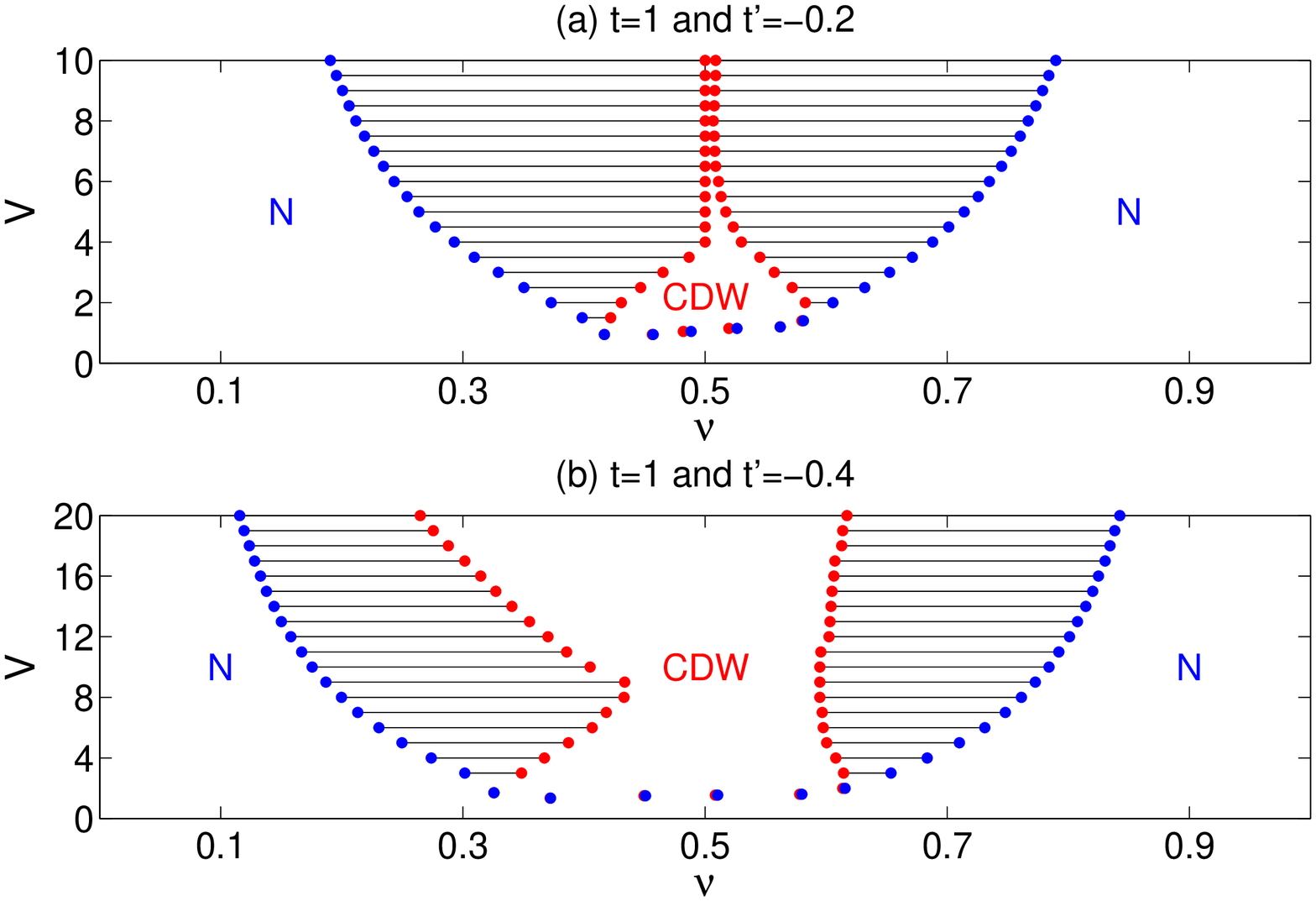}
\end{center}
\caption{Mean field phase diagrams of the 2D \ttpV model at zero
  temperature: coupling $V$ vs.\ filling $\nu$ at $t'=-0.2$ in (a) and
  $t'=-0.4$ in (b) ($t=1$). Shown are the charge-density-wave (CDW),
  normal (N), and mixed (horizontal lines) phases.}
\label{Fig6}
\end{figure}

Figures~\ref{Fig1}(a), \ref{Fig5} and \ref{Fig6} give various phase
diagrams of the 2D \ttpV model. Shown are the phase boundaries of the
CDW phase, the N phase, and the mixed phase (indicated by horizontal
lines) and how they depend on filling $\nu$, coupling $V$ and nnn
hopping $t'$.

As seen in Figure~\ref{Fig1}(a), for $t'=0$ the CDW phase is only
stable at half-filling $\nu=0.5$, and there is a significant mixed
region away from half-filling. The size of this region grows with
increasing $V$.  The invariance of the phase diagram under $\nu\to
1-\nu$ is a consequence of particle-hole symmetry and $t'=0$.  We
found that all phase boundaries in this figure are insensitive to
finite size effects.

The dependence of the phase boundaries on $t'$ and $\nu$ for $V=4$ is
shown in Figure~\ref{Fig5}. Due to particle-hole symmetry this phase
diagram is invariant under $(t',\nu)\to (-t',1-\nu)$, and we therefore
only discuss $t'\leq 0$. For $0\leq -t'< 0.15(3)$, the effect of $t'$
is small and, in particular, the CDW phase exists only at
half-filling. However, for $0.15(3)<- t'<0.21(3)$ it is possible to
dope the CDW phase on the particle (but not on the hole) side, as
discussed in Section~\ref{sec3.1}. For even larger values of $-t'$,
the CDW phase can be doped both on the particle and the hole side, and
the mixed region becomes smaller with increasing $-t'$ and eventually
vanishes.  Note that the phase boundaries between the N and the mixed
phases do not change much with $-t'$, for example,
$\nu_{\N,1}=0.70(1)$ and $0.69(1)$ for $-t'=0$ and $0.3$,
respectively. Moreover, the phase boundary between the CDW and the
mixed phases at the hole side for $-t'>0.21(3)$ is, to a good
approximation, a straight line. The small wiggles of the phase
boundary between the CDW and the mixed phases on the particle side for
$-t'>0.18(3)$ are due to (minor) finite size effects, but all other
phase boundaries are quite insensitive; the same is true for the phase
diagrams in Figure~\ref{Fig6}.

The dependence of the phase boundaries on $t'$ can also be seen in
Figures~\ref{Fig6}(a) and (b) showing the $V$ vs.\ $\nu$ phase
diagrams for $t'=-0.2$ and $-0.4$, respectively. Note that, for
non-zero $t'$, there is a critical coupling value $V_c>0$ below which
no CDW phase exists (e.g.\ $V_c= 0.8(1)$ for $t'=-0.2$), and that
there is a phase boundary between the CDW and N phases.  Moreover, for
strong coupling, the CDW phase broadens out and increasingly dominates
the phase diagram (only visible in (b)). The phase boundary between
the CDW and N phases is quite sensitive to finite size effect, and it
is difficult to determine from our numerical data if it is a first- or
second order phase transition. Finally, to see how the phase diagram
evolves with $t'$ it is instructive to compare Figures~\ref{Fig1}(a)
with Figures~\ref{Fig6}(a) and (b).

\subsection{2D Luttinger model} 
\label{sec4.2}
Apart from Figure~\ref{Fig1}(b), all our phase diagrams for the 2D
Luttinger model have been computed for $t=1$, $V=4$ and $t'=0$ or
$-0.2$. These parameter choices are partly motivated by our results on
the 2D \ttpV model. To be specific, for $t'=-0.2$, the phase diagram
displays several, qualitatively different, features near half-filling,
while $V$ is still of the same order of magnitude as $t$; see
Figure~\ref{Fig6}(a). Furthermore, there is an interesting transition
near $V=4$ and $t'=-0.2$ at which it becomes possible to both
particle- and hole-dope the CDW phase. Other than this, there is
nothing special about these parameter values.

As mentioned in the introduction, a main result of this paper is that
the 2D Luttinger model indeed has a phase in which the antinodal
fermions are gapped and half filled, as conjectured in \cite{EL0,EL1}.
Figure~\ref{Fig1}(b) shows one example with fixed values of $Q$ and
$\kappa$ (as explained later, we can in fact eliminate the parameter
$Q$ in this figure by the requirement $Q=\tQ$).  Similar to the 2D
\ttpV model, we again find a CDW phase, a N phase, and a mixed phase
in-between. However, for $\kappa<1$, the mixed phase is typically much
smaller than for the 2D \ttpV model; cf. Figure~\ref{Fig1}(a).

\begin{figure}[!ht]
   \vspace{0.5in}
\begin{center}
\includegraphics[width=0.9\textwidth]{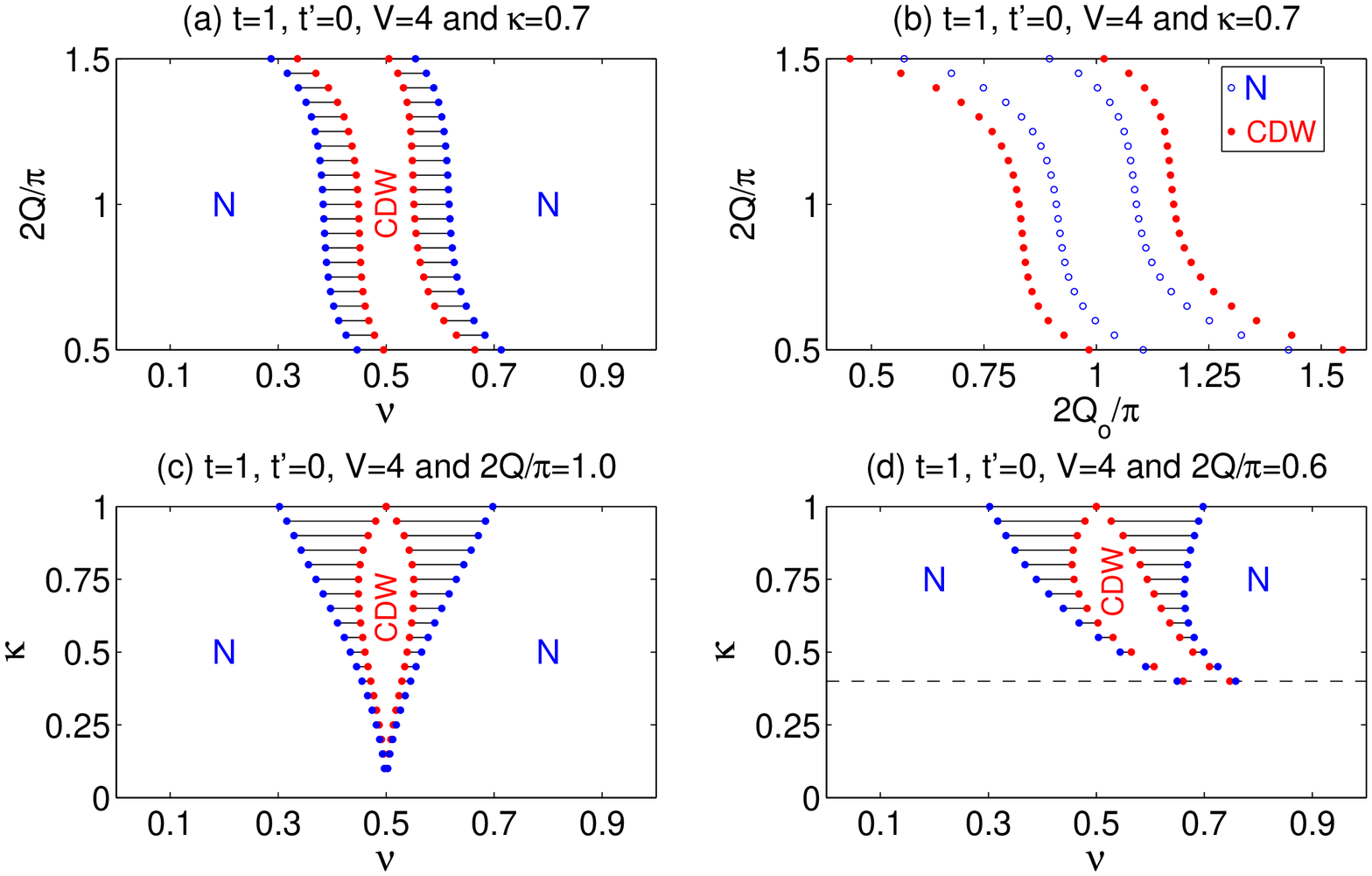}
\end{center}
\caption{Mean field phase diagrams of the 2D Luttinger model at zero
  temperature, $t=1$, $t'=0$, $V=4$, and for different values of the
  antinodal region size $\kappa$ and nodal linearization point $Q$.  Shown
  are the charge-density-wave (CDW), normal (N), and mixed (horizontal
  lines) phase boundaries of the antinodal fermions.  (a) $Q$ vs.\
  total filling $\nu$ at $\kappa=0.7$ (results for other values of
  $\kappa$ are similar). (b) $Q$ vs.\ nodal Fermi surface location
  $\tQ$ at the four phase boundaries in Figure~(a). (c) $\kappa$ vs.\
  $\nu$ at $Q=0.5\pi$. (d) $\kappa$ vs.\ $\nu$ at $Q=0.3\pi$. Note
  that we have to restrict to $\kappa>|1-\frac{2Q}{\pi}|$ as indicated
  by the dashed line.  Due to particle-hole symmetry, the
  corresponding result for $Q=0.7\pi$ can be obtained as $\nu\to
  1-\nu$.}
\label{Fig7}
\end{figure}

An important question is how sensitively the results depend on $Q$ and
$\kappa$.  We find that the qualitative features of the phase diagrams
are robust and the quantitative dependence on $Q$ is weak. However,
the quantitative dependence on $\kappa$ is more pronounced. 
To be more specific, Figures~\ref{Fig7}(a)--(d) give representative
examples for $t'=0$: (a) shows how the phase boundaries depend on $Q$
if $\kappa$ is fixed, while (c) and (d) show how they change with
$\kappa$ for fixed $Q$.  We note that the almost vertical phase
boundaries found in Figure~\ref{Fig7}(a) for $Q$ near $\pi/2$ is not
special to the current choice of $\kappa$ and $t'$.  Particle-hole
symmetry and $t'=0$ imply that the phase boundaries are invariant
under $(Q,\nu)\to (\pi-Q,1-\nu)$. This and $Q=\pi/2$ explain the
symmetry of Figure~\ref{Fig7}(c).  Recall that $\kappa$ is restricted
by \Ref{restrict} to lie in the range $|1-\QQPi|<\kappa<1$.

\begin{figure}[!ht]
   \vspace{0.5in}
\begin{center}
\includegraphics[width=0.7\textwidth]{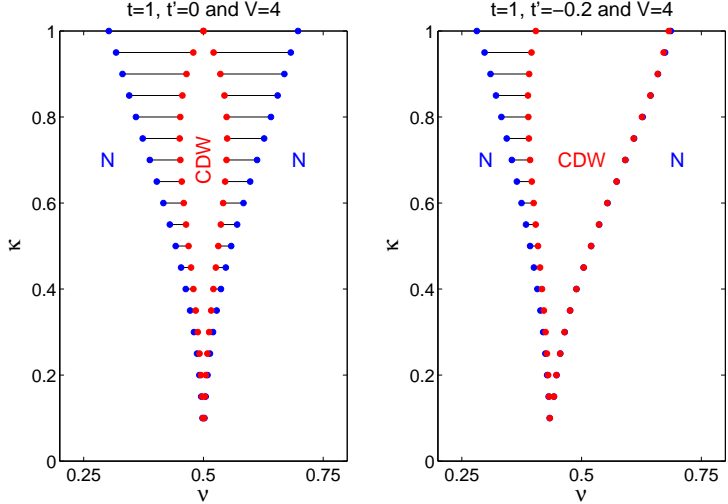}
\end{center}
\caption{Mean field phase diagrams of the 2D Luttinger model at zero
  temperature, $t=1$, $V=4$ and $t'=0$ (left) respectively $t'=-0.2$
  (right): antinodal region size $\kappa$ vs.\ filling $\nu$.  Shown
  are the charge-density-wave (CDW), normal (N), and mixed (horizontal
  lines) phases of the antinodal fermions. The parameter $Q$ is
  determined such that the nodal points are located on the nodal Fermi
  surface arcs at each phase boundary, i.e.\ $Q=\tQ$.}
\label{Fig8}
\end{figure}

Figure~\ref{Fig7}(b) shows $Q$ as a function of the nodal Fermi
surface location, parameterized by $\tQ$, at the four phase boundaries
in \ref{Fig7}(a). These figures suggest that one can fix $Q$ by the
requirement $Q=\tQ$ using the following iterative procedure: given
$Q=Q_n$ one can compute $\tQ$ by solving the mean field equations and
then set $Q_{n+1}=\tQ$.  We generally find that the sequence
$\{Q_n\}_{n=1,2,\ldots}$ converges quickly independent of the starting
value for $Q$.  One can thus eliminate the parameter $Q$ and obtain
phase diagrams depending only on $\kappa$.

Figure~\ref{Fig8} shows two examples of such phase diagrams, the left
for $t'=0$ and the right for $t'=-0.2$.  Comparing the left diagram
with Figure~\ref{Fig7}(c), one finds that the filling values at the
four phase boundaries agree up to an error $\pm 0.002$. Since $Q$
varies over an extended interval in Figure~\ref{Fig8}, while it is
fixed at $\pi/2$ in \ref{Fig7}(c), this further demonstrates the
insensitivity of the phase boundaries to changes in $Q$.  The same
feature holds true for the right diagram. Moreover, Figure~\ref{Fig8}
shows again the qualitative changes of the phase diagram induced by
non-zero $t'$. For $t'=-0.2$, the hole side of the phase diagram is
similar to the one for $t'=0$, but on the particle side, one no longer
finds a mixed region.  Instead there is a continuous transition
between the CDW and the N phase.

\begin{figure}[!ht]
   \vspace{0.5in}
\begin{center}
\includegraphics[width=0.7\textwidth]{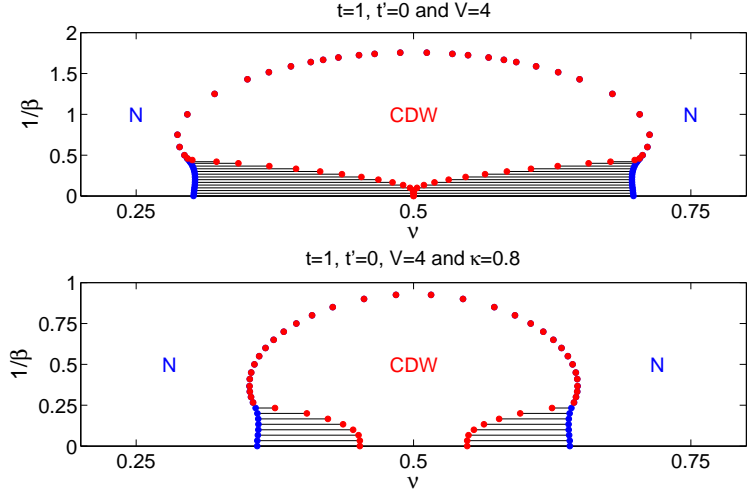}
\end{center}
\caption{Comparison between the temperature ($1/\beta$) vs.\ filling
  ($\nu$) mean field phase diagrams of the 2D \ttpV model (a) and the
  2D Luttinger model (b) for $t=1$, $t'=0$ and $V=4$. In (b),
  $\kappa=0.8$ and $Q$ is fixed by the condition $Q=\tQ$ at the phase
  boundaries. Shown are the charge-density-wave (CDW), normal (N), and
  mixed (horizontal lines) phases. Note the different temperature
  scales in the two plots.}
\label{Fig9}
\end{figure}

Figure~\ref{Fig9} compares the effect of varying the temperature
$1/\beta$ on the phase diagrams of the 2D \ttpV model (a) and the 2D
Luttinger model for $\kappa=0.8$ and $Q=\tQ$ (b). For the 2D \ttpV
model at finite temperature, it becomes possible to dope the CDW phase
away from half-filling. Moreover, the mixed phase decreases in size as
temperature is raised from zero, and it completely disappears at
$1/\beta=0.50(2)$.  For larger values of $1/\beta$, there is a
continuous transition between the CDW and N phase.  The qualitative
features of the 2D Luttinger model at non-zero temperature are quite
similar to the 2D \ttpV model, except that the CDW phase is only
partially gapped and the overall temperature scale is reduced. For
example, the mixed phase now disappears at $1/\beta = 0.25(2)$.  We
note that in computing the filling contribution from the nodal
fermions, we have assumed for simplicity that there are only bosonic
excitations from the ground state, i.e.\ the total number of nodal
fermions is independent of temperature. We leave it to future work to
investigate whether or not this is a justified assumption.

\begin{figure}[!ht]
   \vspace{0.5in}
\begin{center}
\includegraphics[width=0.7\textwidth]{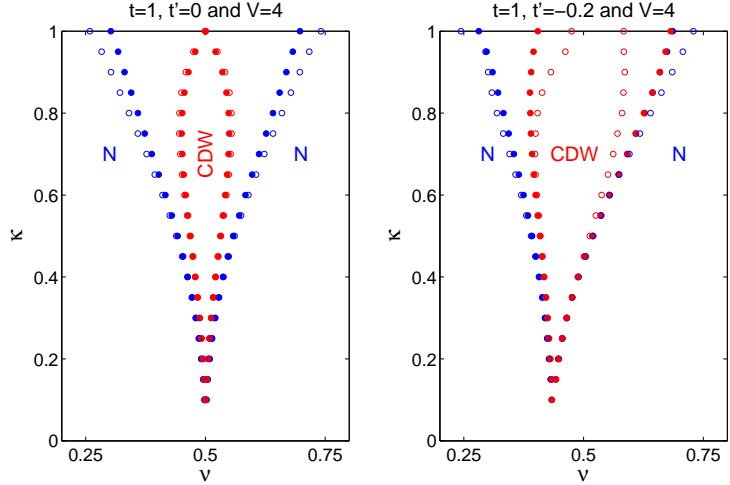}
\end{center}
\caption{Comparison between the results in Figure~\ref{Fig8} obtained
  with the Taylor series approximated band relations \Ref{Epm} (full
  circles), and the ones obtained using the full band relations
  \Ref{Epm0} (open circles).  Note that the results agree well for
  $t'=0$, but there are some differences for $t'=-0.2$ and
  $\kappa>0.6$.}
\label{Fig10}
\end{figure}

Finally, Figure~\ref{Fig10} compares phase diagrams of the 2D
Luttinger model obtained using the exact antinodal band relation in
\Ref{Epm0} (open circles) to the ones obtained using the Taylor series
approximated ones in \Ref{Epm} (full circles).  Note that, for $t'=0$
(left), there are only small quantitative differences. For $t'=-0.2$
(right), however, there are larger deviations, in particular for
$\nu>0.5$ and $\kappa>0.6$: in the former case we find a mixed region
between the CDW- and N phases, but in the latter case this mixed
region is absent.

\subsection{Discussion}
\label{sec4.3}
In this section, we discuss some general features of the mean field
results for the 2D Luttinger- and 2D \ttpV models that we observed in
our numerical computations.  These observations are also based on
phase diagrams not included in the present paper.

Whenever the CDW and the N phase share a phase boundary in a diagram
(i.e. there is no mixed region in between), the CDW order parameter
$\Delta$ goes continuously to zero at the boundary.  Such phase
boundaries are more difficult to determine numerically than
transitions to a mixed phase.

For $t'=0$ we only found a CDW phase in the 2D Luttinger model with
$\nu_a=0.5$, i.e.\ the antinodal fermions are half-filled.  However,
as for the 2D \ttpV model, for non-zero $t'$ it is possible to have a
CDW phase with $\nu_a\neq 0.5$. In fact, as a general rule of thumb,
if the parameters $V$ and $t'$ are such that the CDW phase in the 2D
\ttpV model can be doped, then the 2D Luttinger model has a partially
gapped phase with $\nu_a\neq 0.5$. The converse is not always true. We
expect that the physical properties of the 2D Luttinger model in a
gapped phase with $\nu_a\neq0.5$ is qualitatively different to one
with $\nu_a=0.5$ .

Furthermore, as exemplified in Figure~\ref{Fig10}, when $t'=0$ the
phase diagram of the 2D Luttinger model is hardly changed if one
replaces the Taylor series approximated band relation in \Ref{Epm} by
the exact one in \Ref{Epm0}. For $t'\neq 0$ and $\kappa$ close to 1,
the results are more sensitive to this replacement.

We note that it is not obvious that the phase boundaries can be fixed
unambiguously by the requirement that $Q=\tQ$ since the $\tQ$-value
for which this occurs is, in general, different for the N and the CDW
phase. However, we found that using the $\tQ$-value from the N and the
CDW phase leads to results that are very similar: for $t'=0$ the
discrepancy is typically smaller than the symbol size in our figures,
and this is also true for $t'=-0.2$, apart from the case when $\kappa$
is close to one and the full band relation in \Ref{Epm0} is used. In
Figures~\ref{Fig8}-\ref{Fig10} we determined the phase diagrams using
$\tQ$ from the CDW phase. When using the $\tQ$-values from the N
phase, the CDW phase increases slightly in size, while the N phase
decreases slightly in size.

We now discuss how the size and location of the nodal arcs evolve in
the left diagram of Figure~\ref{Fig8} as $\kappa$ and $\nu$ are
varied. This serves as a representative example for the general
case. Consider first fixed $\kappa$ and $\nu\leq 0.5$ ($\nu\geq 0.5$
is analogous). When the system is in the half-filled and partially
gapped CDW phase, one finds $\tQ=\pi/2$.\footnote{As noted in
  Section~\ref{sec2.2} and further discussed in \cite{EL1}, this
  parameter regime must be interpreted with some care since one has
  additional back-scattering interaction terms not included in the 2D
  Luttinger model when $\tQ=\pi/2$.} As the nodal fermions are hole
doped, $\tQ$ decreases until eventually the CDW phase becomes unstable
at some filling $\nu<0.5$. For even smaller values of $\nu$, the
system is first in a mixed phase, followed by a N phase. When the
system goes from the CDW-mixed phase boundary to the N-mixed boundary,
the $\tQ$ value will first increase towards $\pi/2$. As an example,
when $\kappa=0.8$, $\tQ$ is $0.38(1)$ on the former boundary and
$0.44(1)$ on the latter. Furthermore, when the size of the nodal arcs
is increased (i.e. decreasing the value of $\kappa$), the value of
$\tQ$ on the phase boundaries approaches asymptotically $\pi/2$.

Finally, in mean field theory, the antinodal fermions in the 2D
Luttinger model behave very much as the fermions in the original 2D
\ttpV model scaled by a factor $\kappa^2$. To give an example, when
the antinodal fermions are half-filled ($\nu_a=0.5$), the size of the
CDW gap is, to a good approximation, proportional to
$\kappa^2$. Likewise, the qualitative features of the temperature vs.\
filling phase diagrams for the 2D \ttpV- and 2D Luttinger model are
almost identical if the temperature scale of the latter is reduced by
a factor $\approx\kappa^2$.

\section{Final remarks} 
\label{sec5}
\noindent\textbf{1.}  Mean field theory is a variational method and 
therefore not necessarily restricted to weakly coupled systems. For
example, there exist interacting fermion Hamiltonians of the type
\Ref{Hgen} for which mean field theory is exact (examples include
Hartree- and BCS-like models; see e.g.\ \cite{EL3} for
details). Moreover, for many models describing electrons in
conventional 3D metals, it is known that mean-field type
approximations can give accurate results despite of the presence of
strong Coulomb interactions. Nevertheless, for lattice fermion systems
of Hubbard-type, standard mean field theory fails in a large part of
the parameter regime \cite{LW}.  In this paper, we demonstrated that
it is possible to circumvent this problem by treating parts of the
fermion degrees of freedom exactly using bosonization, as proposed in
\cite{EL0,EL1}. It is straightforward to extend this approach to the
2D Hubbard model \cite{dWL1}.

\noindent\textbf{2.}  The results in \cite{EL1} and the present paper
suggest that the 2D \ttpV model has a qualitatively different behavior
in different filling regimes that can be described by different
effective Hamiltonians. For example, at half filling there is, on the
mean field level, a fully gapped CDW phase that is adequately
described by the antinodal Hamiltonian in \Ref{HLutt} with $\kappa=1$.
Upon doping the system, both the nodal- and the antinodal degrees of
freedom become relevant, and the low-energy physics is governed by the
2D Luttinger model with $\kappa<1$.  For the special case when the
antinodal fermions are gapped, it is possible to describe the system
by a pure nodal fermion model that can be solved exactly by
bosonization \cite{EL1}. For large filling ($\nu$ close to one), the
low-energy physics is expected to be dominated by the out-fermions
with $r=+$ and $s=2$.  In this regime, the appropriate effective model
describes non-interacting fermions with a band relation
$\EE_{+,2}(\vk)\propto (k_1^2+k_2^2)$ \cite{EL1}.  Corresponding
statements hold for the in-fermions at small filling ($\nu$ close to
zero).

\noindent\textbf{3.}  A main result of this paper is that the 2D 
Luttinger model indeed has a partially gapped phase with gapless nodal
fermions and gapped antinodal fermions. The possibility to obtain such
a phase is insensitive to changes of parameter values and model
details like the band relation.

\noindent\textbf{4.} For $\kappa=1$ all degrees of freedom in the 2D
Luttinger model are treated in mean field theory (there are no nodal
fermions then). One should therefore expect that the phase diagram of
the 2D Luttinger model for $\kappa=1$ should be qualitatively similar
to the one of the 2D \ttpV model. Our results show that this is indeed
the case, but there are some quantitative differences; compare
Figures~\ref{Fig5} and \ref{Fig8}. Most of these differences are
explained by the Taylor series approximation of the antinodal bands
\Ref{Epm}; cf.\ Figure~\ref{Fig10}. The remaining discrepancy is due
to the approximation of the interaction vertex mentioned at the end of
Section~\ref{sec2.2}. As discussed there, it is possible to improve on
this approximation \cite{EL1} and derive a refined 2D Luttinger model
that gives back the 2D \ttpV model when setting $\kappa=1$.

\noindent\textbf{5.} There exist parameter values such
that the renormalized antinodal coupling constant $g_a$ in \Ref{ga} is
negative: we found, for example, $g_a<0$ for $t=1$, $t'=-0.2$, $V=10$,
$\kappa=0.8$, and $2Q/\pi=0.24$. However, this and all other such
cases we found are barely within the domain of validity of our method;
recall \Ref{restrict}. We therefore restricted our discussion in this
paper to $g_a>0$.

\noindent\textbf{6.} It is worth noting that our results do not violate 
the Luttinger theorem \cite{Lutt}: the proof of this theorem assumes a
standard connected Fermi surface, and it therefore requires
modifications if there is a (partial) gap.

\noindent\textbf{7.} In this paper, we only presented mean field results 
for the simplest consistent charge-density-wave ansatz (i.e.\ only
$q_0$, $q_1$ and $m_0$ non-zero, see Appendix~\ref{appC}). Another
interesting possibility is the so-called d-wave charge-density-wave
(DDW) phase that has been suggested in the context of high-temperature
superconductors \cite{DDW}. It corresponds to an ansatz with $q_0$,
$q_1$ and $m_2$ non-zero (this is a special case of the general ansatz
in \Ref{MF1}). One can find mean field solutions for which $m_2\neq 0$
gives lower energy than $m_2=0$. However, we never obtained a DDW
solution that has lower energy than the CDW solution discussed in the
main text. The same result hold for the other mean field order
parameters given in Appendix~\ref{appC}.

\noindent \textit{Note added}: Recently, numerical simulations of the
2D \ttpV model using fermionic PEPS \cite{CRBV} produced a phase
diagram that is remarkably similar to our Figure~\ref{Fig1}(a); cf.\
Figure~22 in \cite{CRBV}. 

\bigskip

\subsection*{Acknowledgments} 
This work was supported by the G\"oran Gustafsson Foundation and the
Swedish Research Council.

\appendix
\section{Model details}
\label{appA} 
The results in the present paper are a slight generalization of the
ones derived in \cite {EL1}. In this appendix we explain how these
generalizations are obtained. We also point out the (minor)
differences in notation as compared to \cite {EL1}.  In the following,
we write ``\ELref{33}'' short for ``Equation (33) in \cite{EL1}'' etc.

In \cite{EL1}, the lattice constant $a$ was an important parameter
when taking the partial continuum limit in the nodal region. However,
in the present paper we can without loss of generality set it to $1$.
The relation between the fermion operators used here and the ones in
\cite{EL1} is as follows,
\begin{equation}
\label{ci}
c(\vk)=  \tPiL \hat\psi(\vk),\quad 
c_\pm(\vk)=  \tPiL \hat\psi_{\pm,0}(\vk) . 
\end{equation} 

\subsection{2D \ttpV model}
\label{appA1}
The Hamiltonian in \Ref{H} is obtained from the one in \ELref{19} and
\ELref{25}--\ELref{28} by inserting $\hat u(\vp)=Vu(\vp)/(8\pi^2)$ and
\Ref{ci}, and using
$$
\sum_{\vk_j\in\BZ} v_{\vk_1,\vk_2,\vk_3,\vk_4}
c^\dag(\vk_1) c(\vk_2)c^\dag(\vk_3)c(\vk_4) =
\sum_{\vk_j\in\BZ} v_{\vk_1,\vk_2,\vk_3,\vk_4}
c^\dag(\vk_1)c^\dag(\vk_3)c(\vk_4)c(\vk_2),
$$
i.e.\ normal-ordering of the interaction with respect to the trivial
vacuum does not generate any additional terms.

We recall that the 2D \ttpV model is invariant under the following
transformation of parameters:
\begin{equation}
\label{PH} 
  (t, t',V,\mu,\nu)\to (t, -t',V , 2V - \mu, 1-\nu ).    
\end{equation} 
To see this, we make the particle-hole transformation
$$
  c(\vk)\to c^\dag(-\vk+\vQ),\quad 
c^\dag(\vk)\to c(-\vk+\vQ),\quad
  t'\to-t',\quad \mu\to 2V-\mu  
$$
in \Ref{H} and \Ref{nu} and find
\begin{equation}
  N\to L^2-N,\quad H\to H+(V-\mu)L^2 .  
\end{equation}

\subsection{Effective antinodal Hamiltonian}
\label{appA2}

The 2D \ttpV Hamiltonian \Ref{H} and the effective antinodal
Hamiltonian \Ref{HLutt} are related through a series of steps which we
summarize schematically as
\begin{equation}
\label{schematic}
\HttpV \approx H_n+H_a+H_{na}+\cE_0 \to \Heff +\cE_n+\cE_0 = \HLutt + \cE_a.
\end{equation}
The approximate equality in \Ref{schematic} is the derivation of the
2D Luttinger model from the 2D \ttpV model; see Section~5 in
\cite{EL1}. An important feature in this derivation is normal-ordering
with respect to a non-trivial reference state $\vac$.  This leads to
both constant shifts $\mu\to \mu_{r,s}$ of the chemical potentials for
the different fermion flavors $r=\pm$, $s=0,\pm,2$, and to an additive
energy constant $\cE_0$.  The arrow in \Ref{schematic} involves
integrating out the bosonized nodal fermions in the partition function
for the 2D Luttinger model. This gives an additional energy constant
$\cE_n$ and, after a local-time approximation \cite{EL1}, an effective
Hamiltonian $\Heff$ for (normal-ordered) antinodal fermions. The
normal-ordering is undone on the right-hand side of the equality in
\Ref{schematic}, leading to the antinodal Hamiltonian \Ref{HLutt} and
the energy constant \Ref{cEa}.

The above-mentioned constants play an important role in the present
paper, and we therefore give the details of how to derive them here.
Note that $\EE_{r,0}(\vk)$ in \ELref{47} and \ELref{6} is identical
with $\EE_r(\vk)$ in \Ref {Epm} (recall that $k_\pm=(k_1\pm k_2)/\sqrt
2$).  Moreover, $\Lambda^*_0$ in \ELref{30} and \Ref{BZa} are the
same.

\subsubsection{2D Luttinger model}
In Section~5.3 in \cite{EL1}, the chemical potential $\mu$ was fixed
by the condition that the one-particle states in the nodal regions
$s=\pm$ were completely filled up to the points $\vQ_{r,\pm}=(rQ,\pm
rQ)$. In the present paper the chemical potential is treated as a free
parameter, and the nodal fermions are thus filled up to some points
$(r\tQ,\pm r\tQ)$, with $\tQ$ determined by $\mu$; see \Ref{fixmu}
below. With this generalization, Equations \ELref{59}, \ELref{60}, and
\ELref{9} are modified by the replacement $Q\to\tQ$.

Inserting \ELref{53}, \ELref{54}, and (the modified) \ELref{59} in
\ELref{63}, we obtain for the effective chemical potentials for the
different fermion flavors
\begin{eqnarray}
\label{murs} 
\mu_{\pm,0} &=& \mu - 2V\!\nu - 4t' \nonumber \\
\mu_{\pm,2} &=& \mu - 2V\!\nu +4t' \mp[4t+2VC]\\
\mu_{r,\pm} &=&  \mu - 2V\!\nu +4t\cos(Q)+4t'\cos^2(Q) +2VC\cos(Q) 
\nonumber 
\end{eqnarray} 
with $C$ in \Ref{C} and $\nu$ in \Ref{nu1}.  

As argued in Section~5.3 in \cite{EL1}, the 2D \ttpV Hamiltonian can
be approximated by the normal-ordered 2D Luttinger Hamiltonian without
changing the low-energy physics: $\HttpV\approx H_n+H_a+H_{na}+\cE_0$;
see \ELref{62}. The energy constant appearing here is
\begin{equation}
\label{cE0} 
\cE_0 = \sum_{r=\pm}\sum_{s=0,\pm,2}\cE_{r,s} + \cE_{int} 
\end{equation} 
with 
\begin{equation}
\label{ers} 
\cE_{r,s} = 
\sum_{\vk\in\Lambda^*_{r,s} } (\mbox{$\frac{2\pi}{L}$})^2 
[\eps(\vQ_{r,s}) + \EE_{r,s}(\vk)] 
\langle\mbox{vac}|\hat\psi^\dag_{r,s}(\vk)\hat\psi\pdag_{r,s}(\vk)
|\mbox{vac} \rangle  
\end{equation}
and $\cE_{int}$ in \ELref{65}. The sets $\Lambda^*_{r,s}$ are the
regions of the Brillouin zone corresponding to the $(r,s)$-fermions,
and they are defined in \ELref{42}.  Note that
$\eps(\vQ_{r,s})+\EE_{r,s}(\vk)\approx \eps(\vQ_{r,s}+\vk)$ are the
effective band relations of the $(r,s)$-fermions; see \ELref{46},
\ELref{47} and \ELref{6}. The constants $\cE_{r,s}$ thus correspond to
the expectation values of the kinetic energy of the $(r,s)$-fermions
in the reference state $\vac$. The constant $\cE_{int}$ arises when
the interaction and the term proportional to $-\mu$ are
normal-ordered.

\subsubsection{Effective antinodal model}
As shown in \cite{EL1} Section~6.3, integrating out the nodal fermions
amounts to replacing the 2D Luttinger model by the normal-ordered
effective antinodal Hamiltonian: $H_n+H_a+H_{na}\to \Heff +
\cE_n$. The additional constant is (in the zero temperature limit) the
ground state energy of the nodal Hamiltonian, $\cE_n=\langle
H_n\rangle$.  It can be computed explicitly \cite{EL1}, but in the
present paper we only need its $\mu$-derivative given in \Ref{cEn}.

In the present paper we work with $\HLutt$ in \Ref{HLutt} without
normal-ordering. To find the precise relation between the Hamiltonians
$\Heff$ and $\HLutt$, we use \ELref{37}, \ELref{8} and \ELref{47} to
write \ELref{91} as
\begin{equation}
\begin{split}
  \Heff = \sum_{\vk}(\tPiL)^2\sum_r [\EE_{r,0}(\vk) -\mu_0] 
  :\!\hat\psi\pdag_{r,0}(\vk)\hat\psi^\dag_{r,0}(\vk)\!:\; + \\
  + 2(2V-\geff) \sum_{\vp}(\mbox{$\frac{1}{L}$})^2 
  \hat J_{+,0}(\vp)\hat J_{-,0}(-\vp) . 
\end{split}
\end{equation} 
Inserting $\hat J_{\pm,0}(\vp) = \hat
\rho_{\pm,0}(\vp)-\delta_{\vp,\vzero} L^2\nu_{\pm,0}$, \ELref{46},
\ELref{50}, \Ref{ci} and \ELref{59}, we obtain
$\Heff=\HLutt-\cE_{-,0}-\cE_{+,0} + \cE_\uno$ with $\HLutt$ in
\Ref{HLutt} and the parameters $g_a$ in \Ref{ga}, $\mu_a=\mu_0 +
g_a\nu_a\kappa^2$, $\cE_{\pm,0}$ in \Ref{ers}, and $\cE_\uno$ in
\Ref{cE1}. Inserting \Ref{murs} gives \Ref{mua}.

To compute $\tQ$ we consider the nodal fermion branch $r=s=+$ with the
band relation $\EE_{+,+}(\vk)$ given in \ELref{47} and \ELref{6}. The
condition that these fermions are filled up to $(\tQ,\tQ)$ is
equivalent to $\EE_{+,+}(\vk) -\mu_{+,+} = 0$ for
$\vk+(Q,Q)=(\tQ,\tQ)$, i.e.\
\begin{equation}
\label{fixmu}
\sqrt2 v_F(\tQ-Q) -\mu_{+,+}=0
\end{equation} 
with $v_F$ in \ELref{6}. From this we can compute $\mu$ in
\Ref{fixtQ}, and inserting \Ref{fixtQ} in \Ref{murs} we obtain
($\mu_0\define \mu_{\pm,0}$)
\begin{equation}
\label{mu01}
\mu_0 = \sqrt2 v_F(\tQ-Q) - 4t\cos(Q)- 4t'[1+\cos^2(Q)] -2VC\cos(Q) 
\end{equation} 
which generalizes \ELref{7}. 

\subsubsection{Antinodal energy constant $\cE_a$}
\label{appA2.3}
As discussed above, the two Hamiltonians studied in this paper are
related as $\HttpV\to \HLutt + \cE_a$ with the total energy constant
$\cE_a\define \cE_n+\cE_0-\cE_{+,2}-\cE_{-,2}+\cE_\uno$. Inserting
\Ref{cE0} and using \Ref{murs} we can write this as in \Ref{cEa} with
\begin{equation}
\label{cEkin1}
\cE_{kin} = \cE_{+,2}+\cE_{-,2}+4\cE_{+,+} 
\end{equation} 
where we use $\cE_{+,+}+\cE_{+,-}+\cE_{-,+}+\cE_{-,-}=4\cE_{+,+}$ due
to symmetry. Note that $\cE_{\pm,0}$ in \Ref{cE0} drops out due to undoing the
normal-ordering of the antinodal fermions.
 
In the limit of large $L$, we can compute $\cE_{r,s}/L^2$ as Riemann
integrals (see \Ref{ers}),
\begin{equation}
\iint_{S_{r,s}}\frac{d^2 k}{(2\pi)^2}[\eps(\vQ_{r,s}) + \EE_{r,s}(\vk)],
\end{equation}
with $S_{r,s}\subseteq\Lambda^*_{r,s}$ the set of all occupied states
$\vk$ in $\vac$.  The $(+,2)$-fermions are completely empty, and thus
$\cE_{+,2}=0$.  The $(-,2)$-fermions are completely filled, i.e.\
$S_{-,2}=\Lambda^*_{-2,}$. Using \ELref{42}, \ELref{47} and \ELref{6}
we get
\begin{equation}
\label{cEm2}
\frac{\cE_{-,2}}{L^2} = \int_{-(1-\kappa)\pi/\sqrt2}^{(1-\kappa)\pi/\sqrt{2}}
\frac{dk_+}{2\pi}
\int_{-(1-\kappa)\pi/\sqrt2}^{(1-\kappa)\pi/\sqrt{2}}
\frac{dk_-}{2\pi}[-4(t+t') + (t+2t')(k_+^2+k_-^2)]. 
\end{equation} 
The $(+,+)$-fermions are filled up to $\vk+(Q,Q)=(\tQ,\tQ)$, i.e.\
$S_{+,+}$ contains all $\vk\in\Lambda^*_{+,+}$ with $k_+\leq
\sqrt{2}(\tQ-Q)$. We thus obtain from \ELref{42} and \ELref{47}
\begin{equation}
\label{cEpp}
\frac{\cE_{+,+}}{L^2} =
\int_{-(\kappa\pi+2Q-\pi)/\sqrt{2}}^{\sqrt{2}(\tQ-Q)}\frac{dk_+}{
  2\pi} \int_{-(1-\kappa)\pi/\sqrt2}^{(1-\kappa)\pi/\sqrt{2}}
\frac{dk_-}{2\pi } [-4t\cos(Q)-4t'\cos^2(Q)+v_F k_+]
\end{equation}
with $v_F$ in \ELref{6}.  Computing these integrals we obtain the
result in \Ref{cEkin}. Finally, for $\cE_{int}$ we insert \ELref{53},
\ELref{54} and \ELref{59} in \ELref{65} and obtain by straightforward
but somewhat lengthy computations \Ref{cEint}.

\section{Hartree-Fock theory: Generalities}
\label{appB}
For the convenience of the reader, we collect here the main facts about
Hartree-Fock theory at non-zero temperature. A more detailed account
can be found in \cite{BLS,BP}.

\subsection{Thermal equilibrium states}
\label{appB2}
Many interacting fermion models, including the ones discussed in the
present paper, are given by a Hamiltonian as defined in \Ref{Hgen}.
The restriction to a finite number $\cN$ of one-particle quantum
numbers amounts to having both short- and long distance cutoffs in the
model.

The thermal equilibrium state of \Ref{Hgen} at inverse temperature
$\beta$ is obtained as the minimum over all density matrices $\rho$ of
the grand canonical potential
\begin{equation}
\label{Omega}
\Omega(\rho) \define \textrm{Tr}\bigl(\rho H\bigr) - 
\frac1\beta S(\rho)
\end{equation}
under the particle number constraint
\begin{equation}
  \label{nu_constr} 
  \langle N\rangle  = -\frac{\partial\Omega}{\partial\mu}=N_{0}
\end{equation} 
with ``$\textrm{Tr}$'' the trace in the fermion Fock space, $N_{0}$
some given non-negative integer and
\begin{equation}
S(\rho) \define -\textrm{Tr}\bigl( \rho\ln(\rho)\bigr)
\end{equation}
the entropy of the state.  The (unique) global minimum of $\Omega(\rho)$ is
obtained for the \textit{Gibbs} state
\begin{equation}
  \rho_{\mathrm{Gibbs}}= \frac{\ee^{-\beta H }}{\textrm{Tr}(\ee^{-\beta H})}.  
\end{equation}

\subsection{Unrestricted Hartree-Fock theory}   
\label{appB3}
HF theory at non-zero temperature amounts to restricting the search
for the minimum of $\Omega(\rho)$ to the set of all \textit{HF Gibbs}
states defined as follows,\footnote{All summation indices in this
  section go over $1,2,\ldots,\cN$.}
\begin{equation}
  \rho_{\HFG}= \frac{\ee^{-\beta H_{\HFG} }}{\mathrm{Tr}(\ee^{-\beta H_{\HFG}})}
\end{equation}
with
\begin{equation}
\label{Hqf} 
  H_{\HFG} =\sum_{kl} h\pdag_{kl} c^\dag_k c\pdag_l
\end{equation} 
a Hamiltonian for non-interacting fermions and $h_{kl}$ the matrix
elements of a self-adjoint $\cN\times\cN$ matrix $\mh$.
We refer to $\mh$ as variational one-particle Hamiltonian.

Note that we do not consider the most general definition of a HF
state; one could also allow for states for which $U(1)$ gauge
invariance is broken (see the remark in \ref{appB4}), or states that
are combinations of a pure state and a HF Gibbs state (see \cite{BLS}
for details).

One can compute the grand canonical potential $\OmHF\define
\Omega(\rho_{\HFG})$ for such HF Gibbs states in terms of the
eigenvalues $e_\lambda$ and corresponding orthonormal eigenvectors
$f_\lambda$ of $\mh$ as follows,
\begin{equation}
\label{Omega1}
\begin{split}
  \OmHF = \sum_{kl} (t_{kl}-\mu\delta_{kl})\gamma_{lk} +
  \sum_{klmn}v\pdag_{klmn} 
(\gamma_{mk}\gamma_{nl} -  \gamma_{ml}\gamma_{nk}) 
\\ - \frac1{\beta}\sum_{\lambda}\Bigl(
  \frac{\beta e_\lambda}{\ee^{\beta e_\lambda}+1} + \ln(1+\ee^{-\beta
    e_\lambda}) \Bigr)
\end{split}
\end{equation} 
with
\begin{equation}
\label{one-pdm}
  \gamma_{jk} \define \sum_\lambda \frac1{\ee^{\beta e_\lambda} +1}   
  (f_\lambda)_j \overline{(f_\lambda)_k} 
\end{equation} 
and where $(f_\lambda)_j$ are the components of the eigenvector
$f_\lambda$. Moreover, the particle number constraint is
\begin{equation}
\label{nu_constr1} 
\langle N\rangle = \sum_\lambda  \frac1{\ee^{\beta e_\lambda} +1} =N_0. 
\end{equation} 

In unrestricted HF theory one does not make any restriction on the
variational one-particle Hamiltonian $\mh$, and thus one has, in
principle, $\cN^2$ real variational parameters. However, the number of
variational parameters can often be reduced by the following:

\noindent \textbf{Proposition:} \textit{Local extrema of $\Omega(\rho)$ in
  the set of HF Gibbs states are acquired for a Hamiltonian
\begin{equation}
\label{Hqf1}
H_{\HFG} = \sum_{kl} (t\pdag_{kl}-\mu\delta\pdag_{kl}) 
c^\dag_k c\pdag_l + \sum_{klmn}v\pdag_{klmn}
\bigl( 
c^\dag_kc\pdag_m \langle c^\dag_lc\pdag_n \rangle 
+ \langle c^\dag_kc\pdag_m \rangle c^\dag_lc\pdag_n 
- c^\dag_kc\pdag_n \langle c^\dag_l c\pdag_m \rangle 
- \langle c^\dag_kc\pdag_n \rangle c^\dag_l c\pdag_m 
\bigr) 
\end{equation}
with $\langle c^\dag_k c\pdag_l \rangle=\gamma_{lk}$, i.e.\ when
$\langle\cdot\rangle$ is the expectation value in the state
$\rho_{\HFG}$ corresponding to \Ref{Hqf1}.}

The proof is outlined at the end of this section. The corresponding
result for zero temperature can be found in \cite{BLS}.

It follows that one can restrict the search of the minimum to
variational one-particle Hamiltonians of the form $h_{kl} =
t_{kl}-\mu\delta_{kl}+w_{kl}$, with
\begin{equation}
\label{restrh} 
w\pdag_{kl} =  \sum_{mn} \bigl( 
v\pdag_{knlm} + v\pdag_{nkml} - v\pdag_{knml} - v\pdag_{nklm}
\bigr) \gamma_{mn}
\end{equation} 
for some possible $\gamma_{mn}$ (we will come back to this in
Section~\ref{appB4} below).  For the models of interest to us, this
fact allows to reduce the variational parameters in HF theory
considerably. For example, for the 2D \ttpV model on a square lattice
with $L^2$ sites one would naively expect $L^4$ variational parameters
in unrestricted HF theory, but using \Ref{restrh} this number can be
reduced to $5 L^2$.

Note that \Ref{Hqf1} together with $\langle c^\dag_k c\pdag_l
\rangle=\gamma_{lk}$ gives a self-consistent system of equations that
is often used in practical implementations of HF theory. However, when
restricting HF theory it is important to check that a self-consistent
solution found in this way is indeed an absolute minimum of
\Ref{Omega1} in the set of considered states.  Moreover, in restricted
HF theory it is important to work with the grand canonical ensemble
(i.e.\ fix $\mu$ and not the fermion number) even at zero temperature,
as explained in the main text.

The $\gamma_{jk}$ are the components of a self-adjoint matrix
$\mgamma=(\ee^{\beta\mh}+\id)^{-1}$ called a \textit{one-particle
  density matrix}, and this matrix completely specifies the
corresponding Hartree-Fock Gibbs state. Moreover, the results in
\Ref{nu_constr1} and the second line in \Ref{Omega1} are equivalent to
$\textrm{tr}(\mgamma)=N_0$ and
\begin{equation}
\label{S}
\frac1\beta S(\rho_{\HFG})= \textrm{tr}\bigl(\mh\mgamma +
(1/\beta)\ln(\id+\ee^{-\beta\mh})\bigr) =
-\frac1\beta\textrm{tr}\bigl(\mgamma\ln(\mgamma)+
(\id-\mgamma)\ln(\id-\mgamma)\bigr)
\end{equation} 
with ``$\textrm{tr}$'' the $\cN\times\cN$ matrix trace. Finally, in
the zero-temperature limit $\beta\to\infty$, the one-particle density
matrix \Ref{one-pdm} becomes $\gamma_{jk} = \sum_\lambda
\theta(-e_\lambda)(f_\lambda)_j \overline{(f_\lambda)_k}$; see also
\Ref{one-pdm-zeroT}.

\noindent \textit{Proof of the Proposition:} To find the variation of
the grand canonical potential we use \Ref{Omega1} and \Ref{S}, and we
regard $\mgamma$ rather than $\mh$ as variational parameter:
$\OmHF=\OmHF(\mgamma)$. We compute the variation $\delta\Omega \define
\frac{d}{ds}\Omega_{\HFG}(\mgamma+s\delta\mgamma)|_{s=0}$ and obtain
\begin{equation}
  \delta\OmHF = \sum_{kl}(t_{kl}-\mu\delta_{kl}+w_{kl}-h_{kl} )\delta\gamma_{lk} 
\end{equation}
with $h_{kl}$ the matrix elements of
$\mh=-(1/\beta)\ln[\mgamma(\id-\mgamma)^{-1}]$ and $w_{kl}$ in
\Ref{restrh}; the first three terms are obvious, and the last is
obtained from the variation of the entropy term in \Ref{S}.  This
implies the result. \hfill$\square$

\subsection{Restricted Hartree-Fock theory}
\label{appB4}
The Hamiltonian \Ref{Hgen} often has a large symmetry group, and
assuming that the HF Gibbs state is invariant under (a subgroup of)
these symmetries allows to reduce the number of variational parameters
even further. For example, the 2D \ttpV model is invariant under
translations, parity, and (discrete) rotations. If one restricts to HF
Gibbs states that are invariant under all these transformations, the
number of variational parameters can be reduced to just two; see
Section~\ref{appC1}.  However, as discussed in the main text,
symmetries of the Hamiltonian are often broken, and it is therefore
important to also allow for HF states where (some of) the symmetries
are broken.

In practice, one assumes that the one-particle density matrix commutes
with some suitably chosen subgroup of the original symmetry group.
This leads to restrictions on the matrix elements $\gamma_{kl}$, and
thus, through \Ref{restrh}, on the matrix elements of the variational
one-particle Hamiltonian $\mh$. The latter is true since the
proposition stated in Appendix~\ref{appB3} can be generalized to
restricted HF-theory (we plan to present details elsewhere).

\noindent \textbf{Remark:} We note that the Hamiltonian \Ref{Hgen} is
invariant under the gauge transformation
\begin{equation}
\label{gauge} 
c\pdag_k\to \ee^{\ii \alpha} c\pdag_k
\end{equation}
for real parameters $\alpha$. There exists a natural extension of HF
theory allowing also for states for which this gauge symmetry is
broken.  This extension is relevant for models with attractive
interactions, and its physical interpretation is e.g.
superconductivity.\footnote{There are other physical interpretations
  in the context of nuclear- and elementary particle physics.}
However, since the interaction of the 2D \ttpV model is purely
repulsive, superconducting HF states cannot occur; see \cite{BLS} for
proof.  For the effective antinodal model in \Ref{HLutt} there do
exist parameter regimes where the coupling constant $g_a$ is negative
and thus a superconducting HF state is possible. However, as discussed 
in Section~\ref{sec5}, Remark~5, we found
that this parameter regime is very small. We therefore ignore
superconducting HF states throughout this paper.

\section{Mean field theory} 
\label{appC}
We give here some additional details on our mean field treatments of
the 2D \ttpV- and 2D Luttinger models.

\subsection{2D \ttpV model}
\label{appC1} 
\subsubsection{Symmetries} 
\label{appC1.1}
The 2D \ttpV Hamiltonian in \Ref{H} is invariant under the symmetry
group generated by the following transformations
\begin{equation} 
\label{symm}
\cT_{j}:\; c(\vk)\to \ee^{\ii k_j} c(\vk),\quad
\cP:\; c(\vk)\to c(-\vk),\quad \cR:\; c(\vk)=c(k_1,k_2)
\to c(k_2,-k_1)
\end{equation}
for $j=1,2$. The transformations $\cT_1$ and $\cT_2$, $\cP$ and $\cR$
correspond to translations by the lattice vectors $\ve_1=(1,0)$ and
$\ve_2=(0,1)$, parity transformation, and rotation by $\pi/2$,
respectively.

\subsubsection{Restricted Hartree-Fock theory}
\label{appC1.2}
We consider HF theory restricted to variational states for which the
symmetry is broken down to translations by two sites, i.e.\ the
restricted symmetry group is generated by $(\cT_1)^2$ and
$(\cT_2)^2$. In such a state, one has for the one-particle density matrix
\begin{equation}
\label{MFansatz} 
\gamma(\vk',\vk)=\langle c^\dag(\vk)c(\vk')\rangle = 
\Theta(\vk)\, \delta_{\vk,\vk'} 
+ \tilde\Theta(\vk)\, \delta_{\vk,\vk'+\vQ} 
\end{equation}
with $\vQ=(\pi,\pi)$ and some functions $\Theta$ and $\tilde\Theta$
satisfying
\begin{equation}
\label{conj} 
\Theta(\vk)=\overline{\Theta(\vk)}=\Theta(\vk+2\vQ),
\quad \tilde\Theta(\vk)=
\overline{\tilde\Theta(\vk+\vQ)}= \tilde\Theta(\vk+2\vQ).   
\end{equation} 
Using the fact about HF theory stated in \Ref{Hqf1}\textit{ff} we can
restrict ourselves to HF potentials of the following form,
\begin{equation*}
w(\vk_1,\vk_2) = \frac{2V}{L^2}\sum_{\vk_3,\vk_4\in\BZ} \bigr(
v_{\vk_1,\vk_2,\vk_3,\vk_4} + v_{\vk_3,\vk_4,\vk_1,\vk_2} -
v_{\vk_1,\vk_4,\vk_3,\vk_2} - v_{\vk_3,\vk_2,\vk_1,\vk_4} \bigl)
\gamma(\vk_4,\vk_3). 
\end{equation*} 
Inserting \Ref{MFansatz} and \Ref{v}--\Ref{u}, and using $ u(\vk-\vk')
= \frac12\sum_{j=1}^4 u_j(\vk)u_j(\vk') $ with
\begin{equation}
\label{uj} 
  u_{1,2}\define\cos(k_1)\pm \cos(k_2),\quad 
  u_{3,4}\define\sin(k_1)\pm \sin(k_2)
\end{equation}  
we obtain
\begin{equation}
\label{MF1}
w(\vk,\vk')=  \left(q_0 + \sum_{j=1}^4 q_j u_j(\vk)  \right) 
\, \delta_{\vk,\vk'} + 
\left(m_0 + \sum_{j=1}^4 \ii m_j u_j(\vk)  \right) 
\delta_{\vk,\vk'+\vQ} 
\end{equation} 
with
\begin{equation}
\label{MF_1}
q_0 = 2Vn_0,\quad m_0 = -2V\tilde n_0,
\quad q_j=-\frac{V}2n_j,\quad m_j=-\frac{V}2\tilde n_j
\end{equation} 
and
\begin{equation}
\label{MF_2}
\begin{split} 
  n_0 = \frac1{2L^2}\sum_{\vk}[\Theta(\vk)+\Theta(\vk+\vQ)], \quad
  \tilde{n}_0 =
  \frac1{L^2}\sum_{\vk}\Re\tilde\Theta(\vk) \\
  n_j = \frac1{2L^2}\sum_{\vk}[\Theta(\vk)-\Theta(\vk+\vQ)]u_j(\vk) ,
  \quad \tilde n_j = \frac1{L^2}\sum_{\vk}\Im
  \tilde\Theta(\vk)u_j(\vk)
\end{split} 
\end{equation} 
for $j=1,2,3,4$, with all sums over $\BZ$; we used \Ref{conj} and
$u_j(\vk+\vQ)=-u_j(\vk)$.

\subsubsection{Mean field equations}
\label{appC1.3}
To find $\Theta$ and $\tilde\Theta$ we use Fourier transformation and
write the reference Hamiltonian in \Ref{HR} and \Ref{MF} in matrix
form as follows,
\begin{equation*}
H_{\HFG} = \sum_{\vk\in\BZ_{1/2}} \, 
\left( c^\dag(\vk)\:, \; c^\dag(\vk+\vQ) \right)
\mh(\vk) 
\left(\!\begin{array}{c}c(\vk)\\ c(\vk+\vQ)\end{array}\!\right)
\end{equation*} 
with 
\begin{equation*}
  \mh(\vk) = \left(\begin{array}{cc}
      a_+(\vk)  & b(\vk) 
      \\ \overline{b(\vk)} &
      a_-(\vk)   
\end{array}\right)
\end{equation*}
and
\begin{equation}
\label{ab}  
a_+(\vk)=\epsilon(\vk)+ q_0 -\mu + \sum_{j=1}^4 q_j u_j(\vk),\quad
a_-(\vk)=a_+(\vk+\vQ), 
\quad b(\vk) = m_0 + \ii \sum_{j=1}^4 m_ju_j(\vk) ,
\end{equation} 
where the sum is restricted to half of the Brillouin zone,
$\BZ_{1/2}\define \left\{ \vk\in\BZ:\, k_1>0\right\}$.  
We now can compute
$\mgamma(\vk)\define f(\mh(\vk))$ for $f(x)=1/(\ee^{\beta x}+1)$ using the
following general result,
\begin{equation*}
  f(\mh(\vk)) = \frac{f(e_+(\vk))+f(e_-(\vk))}2 \id +  
  \frac{f(e_+(\vk))-f(e_-(\vk))}{2W(\vk)}\left(\begin{array}{cc}
      a_1(\vk) & b(\vk) 
      \\ \overline{b(\vk)} &
      -a_1(\vk)  
\end{array}\right)
\end{equation*} 
with 
\begin{equation}
\label{epm1} 
e_\pm(\vk) = a_0(\vk) \pm W(\vk),\quad 
W(\vk)=\sqrt{a_1(\vk)^2 + |b(\vk)|^2}, \quad 
a_{0,1}(\vk) =\half[a_+(\vk)\pm a_-(\vk)].
\end{equation}
The $e_\pm(\vk)$ are the eigenvalues of the matrix $\mh(\vk)$ and
equal to the effective mean field band relations. From this and 
\begin{equation*}
\label{TET0} 
\mgamma(\vk) = \left(\begin{array}{cc}
    \Theta(\vk)  & \tilde\Theta(\vk) 
    \\ \overline{\tilde\Theta(\vk)} &
    \Theta(\vk+\vQ)   
\end{array}\right)
\end{equation*} 
we obtain 
\begin{equation}
\label{TET}
\begin{split}
\Theta(\vk) =  \frac12\left( \frac1{\ee^{\beta e_+(\vk)}+1} +
  \frac1{\ee^{\beta e_-(\vk)}+1} \right) +  
\frac{a_1(\vk)}{2W(\vk)}\left( \frac1{\ee^{\beta e_+(\vk)}+1} -
  \frac1{\ee^{\beta e_-(\vk)}+1} \right)\\ 
\tilde\Theta(\vk) =  
\frac{b(\vk)}{2W(\vk)}\left( \frac1{\ee^{\beta e_+(\vk)}+1} -
  \frac1{\ee^{\beta e_-(\vk)}+1} \right). 
\qquad\qquad 
\end{split} 
\end{equation}
By straightforward computations we obtain from \Ref{Omega1}
\begin{equation}
\label{Omega2} 
\begin{split}
  \frac{\OmHF}{L^2} = e_{\HFG} 
  -\sum_{j=0}^4(q\pdag_jn\pdag_j+m\pdag_j\tilde{n}\pdag_j) 
  + V\left(n_0^2 - \tilde n_0^2 -\frac14\sum_{j=1}^4 (n_j^2 +\tilde
    n_j^2) \right)
\end{split} 
\end{equation} 
with
\begin{equation}
\label{eqf} 
e_{\HFG} = - \sum_{r=\pm}\frac1{L^2}\sum_{\vk} 
\frac1{\beta}\ln\bigl(1+\ee^{-\beta
  e_r(\vk)}\bigr), 
\end{equation} 
$n_j$ and $\tilde n_j$ in \Ref{MF_2}, and $e_\pm(\vk)$ in \Ref{epm1}.
Mean field theory amounts to minimizing $\OmHF$ in \Ref{Omega2} with
respect to the 10 variational parameters $q_j$ and $m_j$,
$j=0,1,2,3,4$. Note that, by construction, the saddle point equations $\partial
\OmHF/\partial q_j=\partial \OmHF/\partial m_j=0$ are identical with
the mean field equations in \Ref{MF_1}--\Ref{TET}.

We were only able to find absolute minima of the HF grand canonical
potential with
\begin{equation}
\label{cond} 
  q_2=q_3=q_4=m_2=m_3=m_4=0 . 
\end{equation}
It is interesting to note that \Ref{cond} is equivalent to parity- and
rotation invariance of the HF state: for HF states invariant under
$\cP$ and $\cR$ one has
\begin{equation*}
\label{symm1} 
\Theta(\vk)=\Theta(-\vk)=\Theta(k_2,-k_1)
\end{equation*}
and similarly for $\tilde\Theta$. This implies that $n_j$ and $\tilde
n_j$ in \Ref{MF_2} vanish for $j=2,3,4$, and thus $q_j=m_j$ for
$j=2,3,4$.  Moreover, if this is true, then the HF equations
$q_1=-Vn_1/2$ and $m_1=-V\tilde n_1/2$ are in contradiction unless
$m_1=0$. Note that $m_j=0$ for $j=1,2,3,4$ is equivalent to
\begin{equation*}
\label{m1=0} 
  \overline{\tilde\Theta(\vk)} =\tilde\Theta(\vk). 
\end{equation*} 
We therefore restrict our discussion in the main text to the
simplified ansatz in \Ref{MF} with only three variational parameters
$q_0$, $q_1$ and $m_0 \define \Delta$.

\subsubsection{Numerical details} 
\label{appC1.4}
We solved the mean field equations given above using MATLAB. We found
that it is numerically difficult to minimize the grand canonical
potential $\OmHF$ directly, but, similarly as for the Hubbard model
\cite{BLS}, one can construct an auxiliary extremization problem for a
related potential.  This latter auxiliary potential has
extrema\footnote{By ``extrema'' we mean here points for which all partial
  derivatives are zero.} that coincide with those of $\OmHF$ (in
general, the minima of $\OmHF$ correspond to saddle-points of the
auxiliary potential).  Furthermore, the values of these two potentials
are equal at corresponding extrema.  This auxiliary potential is
numerically well-behaved and it therefore allows to reduce the
computational time considerably (we plan to present details on this
elsewhere).

\subsection{2D Luttinger model}
\label{appC2}
\subsubsection{Symmetries}
\label{appC2.1}
While the effective Hamiltonian in \Ref{HLutt} naturally inherits the
following parity- and rotation symmetries from the 2D \ttpV model,
\begin{equation}
\cP:\;  c_\pm(\vk)\to  c_\pm(-\vk),
\quad 
\cR:\;  c_\pm(\vk)= c_\pm(k_1,k_2) \to  c_\mp(k_2,-k_1),
\end{equation}
its translation invariance is
\begin{equation}
\cT_{\vx}:\; c_\pm(\vk)\to \ee^{\ii \vx\cdot\vk}c_\pm(\vk)
\end{equation}
for all translation vectors $\vx=(x_1,x_2)$ with components
$x_1,x_2\in\mathbb{R}$.  This enhancement of translational symmetry is
due to the continuum limit taken when deriving the 2D Luttinger model,
and it makes it impossible to distinguish between translations by an
even and odd number of lattice sites. We thus have the following
symmetry transformations of $H$ in \Ref{HLutt},
\begin{equation}
  \cC_\alpha:\;  c_\pm(\vk)\to  \ee^{\pm \ii \alpha} c_\pm(\vk)
\end{equation} 
with real $\alpha$. We use particle physics terminology and refer to
$\cC_\alpha$ as chiral transformation.

\subsubsection{Restricted Hartree-Fock theory} 
\label{appC2.2}
It is known that chiral symmetry breaking in a continuum model
corresponds to breaking the translation symmetry from one to two sites
in the corresponding lattice model; see e.g.\ \cite{LS}.  We thus
restrict HF theory to all states invariant under all translations
$\cT_{\vx}$ but not under chiral gauge transformations $\cC_\alpha$,
i.e.\ we make the following ansatz for the one-particle density matrix
\begin{equation}
\label{MFansatzLutt}
\gamma_{r'r}(\vk',\vk)=\langle c^\dag_r(\vk)c\pdag_{r'}(\vk')\rangle = 
\delta_{\vk,\vk'}\Theta_{rr'}(\vk)
\end{equation}
for $r,r'=\pm$ and some functions $\Theta_{rr'}$ obeying
\begin{equation}
\overline{\Theta_{rr'}(\vk)} = \Theta_{r'r}(\vk). 
\end{equation} 
We insert \Ref{MFansatzLutt} in \Ref{restrh} and obtain
\begin{equation*}
H_{\HFG} = H_0 + \frac{2g_a}{L^2}\sum_{\vk,\vk'} \sum_{r,r'=\pm} 
c^\dag_{-r}(\vk)c\pdag_{-r'}(\vk)rr'\Theta_{rr'}(\vk') 
\end{equation*} 
which is equivalent to \Ref{HqfLUTT} with
\begin{equation}
q_0 = g_a n_0,\quad q_1 = -g_a n_1,\quad \Delta = -2g_a\tilde n_0  
\end{equation} 
and 
\begin{equation}
\label{MF_3} 
n_{0,1} = \frac1{L^2}\sum_{\vk} [\Theta_{++}(\vk)\pm \Theta_{--}(\vk)],
\quad \tilde n_0 =  \frac1{L^2}\sum_{\vk} \Theta_{+-}(\vk)
\end{equation} 
with all $\vk$-sums over $\Lambda^*_a$.

\subsubsection{Mean field equations}
\label{appC2.3}
The reference Hamiltonian in \Ref{HqfLUTT} has exactly the same form
as the one treated in Section~\ref{appC1.3}, and we therefore obtain
the same results as in \Ref{epm1} and \Ref{TET} but with
\begin{equation}
\label{abpm} 
a_\pm(\vk) = \EE_\pm(\vk)+q_0 \pm q_1-\mu_a,\quad
b(\vk)=\Delta. 
\end{equation} 
Moreover, as in Section~\ref{appC1.3}
\begin{equation}
\label{Omega3} 
  \frac{\OmHF}{L^2} = e_{\HFG} - q_0 n_0 - q_1 n_1 -
  \Delta\overline{\tilde n_0}-\overline{\Delta}\tilde n_0 + \frac{g_a}2
  \left(n_0^2 - n_1^2 - 4|\tilde n_0|^2\right) + \cE_a
\end{equation} 
with $e_{\HFG}$ as in \Ref{eqf}, $n_{0,1}$ and $\tilde n_1$ in
\Ref{MF_3} and $e_\pm(\vk)$ in \Ref{epm1} and \Ref{abpm}.

\subsection{Consistency check}
\label{appC2.4}
We now show that the condition in \Ref{consistency} holds true.  Using
\Ref{cEkin1}, \Ref{cEm2} and \Ref{cEpp} we get
$$
\frac{\partial(\cE_{kin}/L^2)}{\partial\mu} = [-4t\cos(Q)-4t'\cos^2(Q)
+v_F\sqrt{2}(\tQ -Q)]\frac{\partial}{\partial\mu}(1-\kappa)\tQQPi$$
$$= 
[\mu-2V\!\nu+2VC\cos(Q)]\frac{\partial}{\partial\mu}(\nu-\kappa^2\nu_a) 
$$
where we used \Ref{fixmu}, \Ref{murs} and \Ref{nu1}.  Equation
\Ref{cE1} implies
$$
\frac{\partial(\cE_{\uno}/L^2)}{\partial\mu} =
\nu_a\kappa^2\left(1-2V\frac{\partial\nu}{\partial\mu} +
  g_a\kappa^2\frac{\partial\nu_a}{\partial\mu}\right)+
\left(\mu-2V\!\nu\right)\kappa^2\frac{\partial\nu_a}{\partial\mu}
=\nu_a\kappa^2\frac{\partial\mu_a}{\partial\mu} + (\mu - 2V\!\nu)
\kappa^2\frac{\partial\nu_a}{\partial\mu}
$$
where we used \Ref{mua}. We use \Ref{cEint} and compute, recalling
\Ref{C},
$$
\frac{\partial\cE_{int}}{\partial\mu} = -\nu
-(\mu-2V\!\nu)\frac{\partial\nu}{\partial\mu} -2VC\frac{\partial
  C}{\partial\mu} = -\nu -(\mu-2V\!\nu)\frac{\partial\nu}{\partial\mu}
-2VC\cos(Q)\frac{\partial}{\partial\mu}(\nu-\kappa^2\nu_a). 
$$
We finally use \Ref{cEn}.  Recalling \Ref{cEa} and adding the results
we find that many terms cancel and that the remaining terms add up to
the r.h.s.\ of \Ref{consistency}.


\begin{thebibliography}{99}

\bibitem{EL0} E. Langmann: A two dimensional analogue of the Luttinger
  model, {\tt arXiv:\-math-ph\-/0606041v3} (to appear in
  Lett. Math. Phys.)

\bibitem{EL1} E. Langmann: A 2D Luttinger
  model, {\tt arXiv:0903.0055v3[math-ph]}
  
\bibitem{Bonn} A recent short review is, for example: D. Bonn: Are
  high-temperature superconductors exotic? Nature Physics \textbf{2},
  159 (2006)  

\bibitem{BLS} V.~Bach, E.~Lieb, and J.~Solovej: Generalized
  {H}artree--{F}ock theory and the {H}ubbard model,
  J. Stat. Phys. \textbf{76}, 3 (1994)
  
\bibitem{unrestrHF} J. A. Verges, E. Louis, P. S. Lomdahl, F. Guinea,
  and A. R. Bishop: Holes and magnetic textures in the two-dimensional
  Hubbard model, Phys. Rev. B \textbf{43}, 6099 (1991)
  
\bibitem{LW0} E. Langmann and M. Wallin: Mean-field approach to
  antiferromagnetic domains in the doped Hubbard model, Phys. Rev. B
  \textbf{55}, 9439 (1997)

\bibitem{LW} E. Langmann and M. Wallin: Mean field magnetic phase
  diagrams for the two dimensional $t-t'-U$ Hubbard model,
  J. Stat. Phys. \textbf{127}, 825 (2007)
  
\bibitem{Mattis} D. C. Mattis: Implications of infrared instability in
  a two-dimensional electron gas, Phys. Rev. B \textbf{36}, 745 (1987)

\bibitem{Schulz} H. J. Schulz: Fermi-surface instabilities of a
  generalized two-dimensional {H}ubbard model, Phys. Rev. B
  \textbf{39}, 2940 (1989)

\bibitem{Luther2} A. Luther: Interacting electrons on a square {F}ermi
  surface, Phys. Rev. B \textbf{50}, 11446 (1994)

\bibitem{FRS} N. Furukawa, T. M. Rice, and M. Salmhofer: Truncation of
  a two-dimensional Fermi surface due to quasiparticle gap formation
  at the saddle points, Phys. Rev. Lett. \textbf{81}, 3195 (1998)


\bibitem{CRT} W. R. Czart, S. Robaszkiewicz and B. Tobijaszewska:
  Charge ordering and phase separations in the spinless fermion model
  with repulsive intersite interaction, Acta Phys. Pol. A
  \textbf{114}, 129 (2008)

\bibitem{UV} G. S. Uhrig and R. Vlaming: Inhibition of phase
  separation and appearance of new phases for interacting spinless
  fermions, Phys. Rev. Lett. \textbf{71}, 271 (1993)

\bibitem{KKK} M. Yu. Kagan, K. I. Kugel, and D. I. Khomskii: Phase
  separation in systems with charge ordering, J. Exp. \&
  Theor. Phys. \textbf{93}, 415 (2001)
  
\bibitem{ARPES} For review and further references see: A. Damescelli,
  Z. Hussain and Z.-X. Shen: Angle-resolved photoemission studies of
  the cuprate superconductors, Rev. Mod. Phys. \textbf{75}, 473 (2003)
  
\bibitem{A1} T. Yoshida \textit{et.al.}: Systematic doping evolution of the
  underlying Fermi surface of La$_{2-x}$Sr$_x$CuO$_4$, Phys. Rev. B
  \textbf{74}, 224510 (2006)

\bibitem{A2} K.M. Shen \textit{et.al.}: Nodal quasiparticles and
  antinodal charge ordering in Ca$_{2-x}$\-Na$_{x}$\-CuO$_{2}$\-Cl$_{2}$,
  Science \textbf{307}, 901 (2005)
  
\bibitem{EL3} E. Langmann: Exactly solvable models for 2D interacting
  fermions, J. Phys. A: Math. Gen. \textbf{37}, 407 (2004)
  
\bibitem{dWL1} J. de Woul and E. Langmann, work in progress
  
\bibitem{Lutt} J. M. Luttinger: Fermi surface and some simple
  equilibrium properties of a system of interacting fermions,
  Phys. Rev. \textbf{119}, 1153 (1960)

\bibitem{CRBV} P. Corboz, R. Orus, B. Bauer and G. Vidal: Simulation
  of strongly correlated fermions in two spatial dimensions with
  fermionic Projected Entangled-Pair States, {\tt
    arXiv:0912.0646v1[cond-mat.str-el]}

\bibitem{DDW} S. Chakravarty, R.B. Laughlin, D.K. Morr, and C. Nayak:
  Hidden order in the cuprates, Phys. Rev. B \textbf{63}, 094503
  (2001)

\bibitem{BP} V.~Bach and J.~Poelchau: Hartree-Fock Gibbs states for
  the Hubbard model, Markov Proc.\ and Rel.~Fields \textbf{2}, 225
  (1996)

\bibitem{LS} E. Langmann and G.W. Semenoff: Strong coupling gauge
  theory, quantum spin systems and the spontaneous breaking of chiral
  symmetry, Phys. Lett. B \textbf{297}, 175 (1992)

\end{thebibliography}
\end{document}